
\magnification=\magstephalf

\font\large=cmr17 scaled \magstep0
\font\norm=cmr10 at 10pt
\font\noit=cmti10 at 10pt
\rightline{\norm SPhT 97/078}
\rightline{\tt }
\vskip3cm
\nopagenumbers

  \centerline{\large Generalized Dynkin diagrams and root systems}
\vskip .5truepc
\centerline{\large  and their folding }
\vskip 2truepc
\centerline{\norm Jean-Bernard Zuber}
\bigskip
\centerline{ \noit CEA Saclay, Service de Physique Th\'eorique}
\centerline{ \noit F-91191 Gif sur Yvette Cedex, France}
\vskip2cm
\rm
{\norm \leftskip=2cm \rightskip=2cm \noindent
Graphs which generalize the simple or affine Dynkin diagrams
are introduced. Each diagram defines a bilinear form on  a 
root system and thus a reflection group. We present some
properties of these groups and of their natural ``Coxeter element". 
The folding of these graphs and groups is also discussed, using 
the theory of C-algebras. \par}

\vskip7cm
{\norm Submitted for publication in the proceedings of the Taniguchi Symposium
 {\noit Topological Field Theory, Primitive Forms and Related Topics}, 
 Kyoto Dec 1996, M. Kashiwara, A. Matsuo,  K. Saito and I. Satake eds, 
 Birkha\"user.}
\vfill\eject


\input amssym.def
\input amssym.tex
\font\elevenrm=cmr10 at 11pt
\font\it=cmti10 at 11pt
\font\bf=cmbx10 at 11pt
\font\rm=cmr10 at 11pt
\font\sl=cmsl10 at 11pt

\catcode`\@=12

\voffset=2.5truepc
\hoffset=2.5truepc
\hsize=5.15truein
\vsize=8.5truein 

\font\ninerm=cmr9 at 9pt
\font\ninebf=cmbx10 at 9pt
\font\nineit=cmti9 at 9pt

\tolerance=6000
\parskip=0pt
\parindent=18pt
\font\elevenrm=cmr10 at 11pt
\font\elevenit=cmti10 at 11pt

 \abovedisplayskip=12pt plus3pt minus2pt
 \belowdisplayskip=12pt plus3pt minus2pt
 \abovedisplayshortskip =12pt plus3pt minus2pt
 \belowdisplayshortskip =12pt plus3pt minus2pt

 \def\lskipamount{12pt}
 \def\lskip{\vskip\lskipamount plus3pt minus2pt}
 \def\lbreak{\par \ifdim\lastskip<\lskipamount
  \removelastskip \penalty-200 \lskip \fi}

 \def\lnobreak{\par \ifdim\lastskip<\lskipamount
  \removelastskip \penalty200 \lskip \fi}

\baselineskip=12.8pt

\font\titlefont=cmbx12 at 14pt
\font\sectionfont=cmbx12 at 11pt
\font\sevenpoint=cmr7 at 7pt

\font\elevenbf=cmbx10 at 11pt
\font\elevenit=cmti10 at 11pt

\font\autit=cmti12 

\def\auth#1{{\autit {\centerline{#1}}}\vskip 3truepc}

\def\ResHead#1{\vskip 1.5truepc\centerline{\hbox {{\sectionfont
#1}}}\vskip1truepc\noindent}
\def\AHead#1{\vskip 1.5truepc\centerline{\hbox {{\sectionfont
#1}}}\vskip1truepc\noindent}

\def\RefHead#1{\vskip 1truepc\centerline{\hbox {{\sectionfont
#1}}}\vskip4pt\noindent}

\def\ref#1{\vskip 1.5pc{\centerline {\elevenbf  References}}\vskip
4pt \noindent}
\def\affil#1{\vskip 1truepc{\noindent {\ninerm #1}}}

\headline={\ifodd\pageno\rightheadline \else
    \leftheadline\fi}
 \def\rightheadline{\elevenrm\hfill{\elevenit {}
}\hfil\elevenrm\folio}
\def\leftheadline{\elevenrm\folio\hfill{\elevenit {}}}
\footline={\hfill}


\font\elevenrm=cmr10 at 11pt
\font\it=cmti10 at 11pt
\font\bf=cmbx10 at 11pt

\font\rm=cmr10 at 11pt
\font\rmn=cmr9

\pageno=1
\overfullrule=0pt

  \centerline{\titlefont Generalized Dynkin diagrams and root systems}
\vskip .5truepc
\centerline{\titlefont  and their folding }
\vskip 2truepc
\auth{Jean-Bernard Zuber}

\rm
\ResHead{Abstract}  
Graphs which generalize the simple or affine Dynkin diagrams
are introduced. Each diagram defines a bilinear form on  a 
root system and thus a reflection group. We present some
properties of these groups and of their natural ``Coxeter element". 
The folding of these graphs and groups is also discussed, using 
the theory of C-algebras.



%
\def\CA{{\cal A}}	\def\CB{{\cal B}}	\def\CC{{\cal C}}
\def\CD{{\cal D}}	\def\CE{{\cal E}}	
\def\CG{{\cal G}}		\def\CI{{\cal J}}
		
	\def\CN{{\cal N}}	
\def\CP{{\cal P}}		\def\CR{{\cal R}}
		
\def\CV{{\cal V}}		

\def\({ \left( }\def\[{ \left[ }
\def\){ \right) }\def\]{ \right] }
%


\def\IR{\relax{\rm I\kern-.18em R}}
\font\cmss=cmss10 \font\cmsss=cmss10 at 7pt
\def\IZ{\relax\ifmmode\mathchoice
{\hbox{\cmss Z\kern-.4em Z}}{\hbox{\cmss Z\kern-.4em Z}}
{\lower.9pt\hbox{\cmsss Z\kern-.4em Z}}
{\lower1.2pt\hbox{\cmsss Z\kern-.4em Z}}\else{\cmss Z\kern-.4em Z}\fi}
\def\inbar{\,\vrule height1.5ex width.4pt depth0pt}
\def\IB{\relax{\rm I\kern-.18em B}}
\def\IC{\relax\hbox{$\inbar\kern-.3em{\rm C}$}}
\def\ID{\relax{\rm I\kern-.18em D}}
\def\IE{\relax{\rm I\kern-.18em E}}
\def\IF{\relax{\rm I\kern-.18em F}}
\def\IG{\relax\hbox{$\inbar\kern-.3em{\rm G}$}}
\def\IH{\relax{\rm I\kern-.18em H}}
\def\II{\relax{\rm I\kern-.18em I}}
\def\IK{\relax{\rm I\kern-.18em K}}
\def\IL{\relax{\rm I\kern-.18em L}}
\def\IM{\relax{\rm I\kern-.18em M}}
\def\IN{\relax{\rm I\kern-.18em N}}
\def\IO{\relax\hbox{$\inbar\kern-.3em{\rm O}$}}
\def\IP{\relax{\rm I\kern-.18em P}}
\def\IQ{\relax\hbox{$\inbar\kern-.3em{\rm Q}$}}
\def\IGa{\relax\hbox{${\rm I}\kern-.18em\Gamma$}}
\def\IPi{\relax\hbox{${\rm I}\kern-.18em\Pi$}}
\def\ITh{\relax\hbox{$\inbar\kern-.3em\Theta$}}
\def\IOm{\relax\hbox{$\inbar\kern-3.00pt\Omega$}}
\def\N{\IN}


\def\IC{\relax\hbox{$\inbar\kern-.3em{\rm C}$}}

\input amssym.def
\input amssym.tex
\def\IZ{\Bbb Z}\def\IR{\Bbb R}\def\IC{\Bbb C}\def\IN{\Bbb N}





\def\oh{{1\over 2}}\def\un{{\bf 1}}

\def\Ga{\alpha}\def\Gb{\beta}\def\Gc{\gamma}\def\GC{\Gamma}
\def\Gd{\delta}\def\GD{\Delta}\def\Ge{\epsilon}

\def\Gl{\lambda}\def\GL{\Lambda}
\def\Gm{\mu}\def\Gn{\nu}
\def\Gr{\rho}
\def\Gs{\sigma}\def\GS{\Sigma}\def\Gt{\tau}

\def\Go{\omega}

\def\mod{{\hbox { mod}}\, }
\def\modmod{{\hbox{\sevenpoint mod}}\, }
\def\det{{\hbox { det}}\, }
\def\diag{\hbox{ diag }}
 
\def\bra{\langle}\def\ket{\rangle}
\def\rep{representation}  
\def\exp{\hbox{exp}\,} 
\def\cos{\hbox{ cos}} \def\sin{\hbox{ sin}} 
%


\def\nind{{\par\noindent}}
\def\slh{\widehat{sl}}
\def\un{{\hbox {\bf 1}} }
\def\II{{\hbox {I\hskip-2pt I}}}
\def\Exp{\hbox{Exp}}
\def\Expexp{\hbox{\sevenpoint Exp}}

\def\o{{\aleph}}
\def\o{1\!\!\!\circ}

\def\hA{\hat A}\def\hD{\hat D}\def\hE{\hat E}\def\hT{\hat T}
\def\hB{\hat B}\def\hC{\hat C}\def\hF{\hat F}\def\hG{\hat G}
\def\ba{\bar a}
\def\ve{\varepsilon}

\def\tvp{\vrule height 2pt depth 1pt} 
\def\thp{\vrule height 0.4pt width 0.35em}
\def\cc#1{\hfill#1\hfill}
\def\nolabels{\def\wrlabeL##1{}\def\eqlabeL##1{}\def\reflabeL##1{}}
\def\writelabels{\def\wrlabeL##1{\leavevmode\vadjust{\rlap{\smash%
{\line{{\escapechar=` \hfill\rlap{\sevenrm\hskip.03in\string##1}}}}}}}%
\def\eqlabeL##1{{\escapechar-1\rlap{\sevenrm\hskip.05in\string##1}}}%
\def\reflabeL##1{\noexpand\llap{\noexpand\sevenrm\string\string\string##1}}}
\nolabels
%
\global\newcount\secno \global\secno=0
\global\newcount\meqno \global\meqno=1
\def\newsec#1{\global\advance\secno by1\message{(\the\secno. #1)}
\global\subsecno=0\eqnres@t\noindent{\bf\the\secno. #1}
\writetoca{{\secsym} {#1}}\par\nobreak\medskip\nobreak}
\def\eqnres@t{\xdef\secsym{\the\secno.}\global\meqno=1\bigbreak\bigskip}
\def\sequentialequations{\def\eqnres@t{\bigbreak}}\xdef\secsym{}
\global\newcount\subsecno \global\subsecno=0
\def\subsec#1{\global\advance\subsecno by1\message{(\secsym\the\subsecno. #1)}
\ifnum\lastpenalty>9000\else\bigbreak\fi
\noindent{\it\secsym\the\subsecno. #1}\writetoca{\string\quad
{\secsym\the\subsecno.} {#1}}\par\nobreak\medskip\nobreak}
\def\appendix#1#2{\global\meqno=1\global\subsecno=0\xdef\secsym{\hbox{#1.}}
\bigbreak\bigskip\noindent{\bf Appendix #1. #2}\message{(#1. #2)}
\writetoca{Appendix {#1.} {#2}}\par\nobreak\medskip\nobreak}
%
%
\def\eqnn#1{\xdef #1{(\secsym\the\meqno)}\writedef{#1\leftbracket#1}%
\global\advance\meqno by1\wrlabeL#1}
\def\eqna#1{\xdef #1##1{\hbox{$(\secsym\the\meqno##1)$}}
\writedef{#1\numbersign1\leftbracket#1{\numbersign1}}%
\global\advance\meqno by1\wrlabeL{#1$\{\}$}}
\def\eqn#1#2{\xdef #1{(\secsym\the\meqno)}\writedef{#1\leftbracket#1}%
\global\advance\meqno by1$$#2\eqno#1\eqlabeL#1$$}
%
\newskip\footskip\footskip14pt plus 1pt minus 1pt 
\def\footnotefont{\ninepoint}\def\f@t#1{\footnotefont #1\@foot}
\def\f@@t{\baselineskip\footskip\bgroup\footnotefont\aftergroup\@foot\let\next}
\setbox\strutbox=\hbox{\vrule height9.5pt depth4.5pt width0pt}
\global\newcount\ftno \global\ftno=0
\def\foot{\global\advance\ftno by1\footnote{$^{\the\ftno}$}}
%
\newwrite\ftfile
\def\footend{\def\foot{\global\advance\ftno by1\chardef\wfile=\ftfile
$^{\the\ftno}$\ifnum\ftno=1\immediate\openout\ftfile=foots.tmp\fi%
\immediate\write\ftfile{\noexpand\smallskip%
\noexpand\item{f\the\ftno:\ }\pctsign}\findarg}%
\def\footatend{\vfill\eject\immediate\closeout\ftfile{\parindent=20pt
\centerline{\bf Footnotes}\nobreak\bigskip\input foots.tmp }}}
\def\footatend{}
%
%
\global\newcount\refno \global\refno=1
\newwrite\rfile
\def\ref{[\the\refno]\nref}
\def\nref#1{\xdef#1{[\the\refno]}\writedef{#1\leftbracket#1}%
\ifnum\refno=1\immediate\openout\rfile=refs.tmp\fi
\global\advance\refno by1\chardef\wfile=\rfile\immediate
\write\rfile{\noexpand\item{#1\ }\reflabeL{#1\hskip.31in}\pctsign}\findarg}
\def\findarg#1#{\begingroup\obeylines\newlinechar=`\^^M\pass@rg}
{\obeylines\gdef\pass@rg#1{\writ@line\relax #1^^M\hbox{}^^M}%
\gdef\writ@line#1^^M{\expandafter\toks0\expandafter{\striprel@x #1}%
\edef\next{\the\toks0}\ifx\next\em@rk\let\next=\endgroup\else\ifx\next\empty%
\else\immediate\write\wfile{\the\toks0}\fi\let\next=\writ@line\fi\next\relax}}
\def\striprel@x#1{} \def\em@rk{\hbox{}}
\def\lref{\begingroup\obeylines\lr@f}
\def\lr@f#1#2{\gdef#1{\ref#1{#2}}\endgroup\unskip}

\def\addref#1{\immediate\write\rfile{\noexpand\item{}#1}} 
\def\startrefs#1{\immediate\openout\rfile=refs.tmp\refno=#1}
\def\xref{\expandafter\xr@f}\def\xr@f[#1]{#1}
\def\refs#1{\count255=1[\r@fs #1{\hbox{}}]}
\def\r@fs#1{\ifx\und@fined#1\message{reflabel \string#1 is undefined.}%
\nref#1{need to supply reference \string#1.}\fi%
\vphantom{\hphantom{#1}}\edef\next{#1}\ifx\next\em@rk\def\next{}%
\else\ifx\next#1\ifodd\count255\relax\xref#1\count255=0\fi%
\else#1\count255=1\fi\let\next=\r@fs\fi\next}
\newwrite\ffile\global\newcount\figno \global\figno=1
\def\fig{fig.~\the\figno\nfig}
\def\nfig#1{\xdef#1{fig.~\the\figno}%
\writedef{#1\leftbracket fig.\noexpand~\the\figno}%
\ifnum\figno=1\immediate\openout\ffile=figs.tmp\fi\chardef\wfile=\ffile%
\immediate\write\ffile{\noexpand\medskip\noexpand\item{Fig.\ \the\figno. }
\reflabeL{#1\hskip.55in}\pctsign}\global\advance\figno by1\findarg}
\def\xfig{\expandafter\xf@g}\def\xf@g fig.\penalty\@M\ {}
\def\figs#1{figs.~\f@gs #1{\hbox{}}}
\def\f@gs#1{\edef\next{#1}\ifx\next\em@rk\def\next{}\else
\ifx\next#1\xfig #1\else#1\fi\let\next=\f@gs\fi\next}
\newwrite\lfile
{\escapechar-1\xdef\pctsign{\string\%}\xdef\leftbracket{\string\{}
\xdef\rightbracket{\string\}}\xdef\numbersign{\string\#}}

\def\writestop{\def\writestoppt{\immediate\write\lfile{\string\pageno%
\the\pageno\string\startrefs\leftbracket\the\refno\rightbracket%
\string\def\string\secsym\leftbracket\secsym\rightbracket%
\string\secno\the\secno\string\meqno\the\meqno}\immediate\closeout\lfile}}
\def\writestoppt{}\def\writedef#1{}
\def\seclab#1{\xdef #1{\the\secno}\writedef{#1\leftbracket#1}\wrlabeL{#1=#1}}
\def\subseclab#1{\xdef #1{\secsym\the\subsecno}%
\writedef{#1\leftbracket#1}\wrlabeL{#1=#1}}
\newwrite\tfile \def\writetoca#1{}
\def\leaderfill{\leaders\hbox to 1em{\hss.\hss}\hfill}
\def\writetoc{\immediate\openout\tfile=toc.tmp
   \def\writetoca##1{{\edef\next{\write\tfile{\noindent ##1
   \string\leaderfill {\noexpand\number\pageno} \par}}\next}}}

%
\def\subsec#1{\nind{\it #1}\smallskip\nind }

\input epsf.tex
\newcount\figno
\figno=0
\def\fig#1#2#3{
\par\begingroup\parindent=0pt\leftskip=1cm\rightskip=1cm\parindent=0pt
\baselineskip=11pt
\global\advance\figno by 1
\midinsert
\epsfxsize=#3    
\centerline{\epsfbox{#2}}
\vskip 12pt
{\ninebf Fig. \the\figno:} #1\par
\endinsert\endgroup\par
}
\def\figlabel#1{\xdef#1{\the\figno}}
\def\encadremath#1{\vbox{\hrule\hbox{\vrule\kern8pt\vbox{\kern8pt
\hbox{$\displaystyle #1$}\kern8pt}
\kern8pt\vrule}\hrule}}

\AHead{Introduction}
This paper is devoted to the study of generalized Dynkin diagrams, 
regarded as encoding the geometry of a root system, and of the 
ensuing reflection groups. The  graphs
are $N$-colourable  --in a sense to be defined
below-- while the ordinary simple or affine Dynkin diagrams 
correspond to the case $N=2$. In fact, the main features of the 
construction are based on the algebra $sl(N)$ or its affine 
extension $\widehat{sl}(N)$, with the ordinary (simple or affine)
 Dynkin diagrams and 
finite or affine reflection groups being associated with $sl(2)$. 
Ordinary $ADE$ (resp affine $\hA\hD\hE$) Dynkin diagrams are 
known to be the only bi-colourable graphs with a spectrum of their
adjacency matrix less than (resp. less than or equal to)  2. Here we use
an appropriate generalization of this spectral property.

This generalization has emerged in a natural way in several 
related problems of current interest in mathematical physics: 
conformal field theories, integrable lattice models, 
topological field theories (for a review, see for example 
[Zi]) and  3-manifold invariants [O].

In section  1 we introduce 
the notations and the axioms imposed on the graphs. 
Section 2 presents the main results on the associated reflection 
groups: signature of the metric, definition of a natural Coxeter 
element and its order.  
In section 3 is discussed the folding of ordinary and 
generalized Dynkin diagrams. While the folding of simply laced
into non simply laced Dynkin diagrams, for example 
$E_8$ into $H_4$, is usually found in some empirical way [Y,Sh,MP], 
it is shown 
that there is a general way to find such possible foldings, based on the
structure of the so-called ``C-algebras" [BI] attached to the graph 
(for a review, see for example [Zr]).  
Some aspects of the connection with singularities are discussed in sect. 4. 
as well as some indications about the connections with 
superconformal and topological field theories. 

This paper is an extension of a former exposition [Z] but may be read
independently. The main 
new results concern: i) the extension of the discussion from graphs
generalizing the ordinary Dynkin diagrams of finite type 
to a larger class encompassing the graphs of affine type; 
ii) the discussion of the folding of (generalized) Dynkin diagrams
and reflection groups.  

\AHead{1. Generalized Dynkin diagrams}
%
\subsec{1.1 Notations on $sl(N)$} 
We shall consider a class of graphs which generalize the classical $A,D,E$ 
and extended $\hat A, \hat D, \hat E$ Dynkin diagrams. While the $ADE$
or $\hA\hD\hE$ diagrams may
be regarded as related to the $sl(2)$ algebra, the new ones are associated 
with $sl(N)$. 
Let $\GL_1, \cdots, \GL_{N-1}$ be the fundamental weights of $sl(N)$. 
 Let $\Gr=\GL_1+\cdots + \GL_{N-1}$ denote their sum.
I denote the weight lattice by $\CP$. Let $k\in \IN$.
I recall that the set of integrable weights (shifted by $\Gr$) of the 
affine algebra $\slh(N)$ at level $k$ is the following subset of $\CP$ 
\eqn\Ia{
\CP^{(h)}_{++}= \{\Gl=\Gl_1 \GL_1 +\cdots +\Gl_{N-1}\GL_{N-1}
|\Gl_i\in \IN,\,  \Gl_i \ge 1,\, \Gl_1+\cdots + \Gl_{N-1}\le h-1 \} \ ,}
where $h=k+N$. We shall also consider the larger set 
\eqn\Ib{
\CP^{(h)}_+= \{\Gl=\Gl_1 \GL_1 +\cdots +\Gl_{N-1}\GL_{N-1}
  |\Gl_i\in \IN,\, \Gl_i \ge 0, \Gl_1+\cdots + \Gl_{N-1}\le h \}\ .}
These two sets admit two natural automorphisms: 
\item{} the conjugation $\CC$ of representations
\eqn\Ic{
\CC : \quad \Gl=(\Gl_1, \Gl_2,\cdots , \Gl_{N-1} ) \mapsto
\bar \Gl = ( \Gl_{N-1},\cdots , \Gl_2,  \Gl_1)}
%
\item{} and the $\IZ_N$ automorphism
%
\eqn\Id{
\Gs  : \quad \Gl=(\Gl_1, \Gl_2,\cdots , \Gl_{N-1} ) \mapsto
\ \Gl'=(h- \sum_{j=1}^{N-1} \Gl_j, \Gl_1, \Gl_2,\cdots , \Gl_{N-2} )\ .}

\nind
We then introduce the weights $e_i$ of the standard $N$-dimensional
representation of $sl(N)$
\eqn\Ie{
e_1 =\GL_1, \quad e_i=\GL_i-\GL_{i-1},(i=2,\cdots,N-1),\quad e_N=-\GL_{N-1} }
endowed with the scalar product $(e_i, e_j)=\Gd_{ij}- {1\over N}$. 

%
\setbox34=\hbox{$\scriptstyle {p}$} %
\setbox3=\hbox{$\vcenter{\offinterlineskip
\+ \thp&\cr  
\+ \tvp\cc{}&\tvp\cr 
\+ \thp&\cr  
\+ $\!{}^{\vdots}$\cc{}&$\!{}^{\vdots}$\cr 
\+ \thp&\cr  
\+ \tvp\cc{}&\tvp\cr 
\+ \thp&\cr  }$}
\setbox22=\hbox{$\left.\vbox to \ht3{}\right\}$} 
%
%
\font\elevenrm=cmr10 at 11pt
\headline={\ifodd\pageno\rightheadline \else
    \leftheadline\fi}
\def\rightheadline{{\elevenit Generalized Dynkin diagrams}
\hfill \elevenrm\folio}
    \def\leftheadline{\elevenrm \folio \hfill{\elevenit {
J.-B. Zuber}}}
 \footline={\hfill}
\bigskip
\bigskip
\subsec{1.2 Axioms on the graphs} %
\nind
The graphs are defined by the following data and axioms
\item{1)} A set $\CV$ of $|\CV|=n$ vertices is given. These vertices
 are denoted by latin letters $a,b, \cdots$ 
There exists an involution $a \mapsto \bar a$ and the 
set $\CV$ admits a $\IZ_N$ grading denoted $\tau(a)$ and called
$N$-ality or color, such that $\tau(\bar a)=-\tau(a) \mod  N$. 
\item{2)} A set of $N-1$ commuting $n\times n$ matrices $G_p$ is given. They 
may be thought of as labelled by the fundamental representations of $sl(N)$, 
$G_p=G_{\copy3\copy22\copy34}$, $ 1\le p\le N-1$. 
Their matrix elements are assumed to be non negative integers, 
so that they may be regarded as adjacency matrices of $N-1$ graphs
$\CG_p$. Multiple edges $(G_p)_{ab}\ge 2$ are allowed. 
(The integrality assumption may be relaxed to allow the analogue 
of non simply-laced Coxeter-Dynkin diagrams, see below Remark 2.) 
We also assume that the graph  $\CG_1$ is connected.
\item{3)} The edges of the graphs $\CG_p$ are compatible with the grading 
$\Gt$ in the sense that 
\eqn\If{
(G_p)_{ab} =0 \quad \hbox{ if } \quad \Gt(b)\ne \Gt(a)+p \ \hbox{ mod } N \ .}
Thus for $p\ne {N\over 2}$, the edges are oriented. Also note that for 
a given pair $(a,b)$, there is at most one graph $\CG_p$ 
with an edge between $a$ and $b$. 
\item{4)} The matrices are transposed of one another
\eqn\Ig{
G^t_p= G_{N-p} }
and are invariant under the involution $a \mapsto \bar a$ in the sense that
%
\eqn\Ih{
(G_p)_{\bar a \bar b} =   (G_p)_{ba}\ .}
\item{5)} As a consequence of 2) and 4), 
the matrices $G_p$ commute with their transpose (``normal 
matrices'') 
and may thus be simultaneously diagonalized in a common orthonormal 
basis. This basis,  denoted $\psi^{(\Gl)}$, is {\it assumed} to be 
labelled by weights $\Gl$ of $sl(N)$, that are  restricted to be
in $\CP^{(h)}_{+}$, 
--or even more restricted to be in $\CP^{(h)}_{++}$ (see below)--
, for some integer $h \ge N$, in such a way that 
the eigenvalues $\Gc_p^{(\Gl)}$ have the form 
\eqn\Ii{
\Gc_p^{(\Gl)} = \chi_p(M(\Gl)) }
where $\chi_p$ is the ordinary character for the $p$-th fundamental
representation of the group $SU(N)$, and $M(\Gl)$ denotes the 
diagonal matrix $M(\Gl)= \diag(\ve_j(\Gl))_{j=1,
 \cdots N}$. Here and throughout this article,
\eqn\Iia{ \ve_j(\Gl):=   \exp -{2i\pi\over h} (e_j,\Gl)\ .}
Maybe more explicitly, the $\Gc_p$ read 
\eqn\Ij{
\eqalign{\gamma^{(\Gl)}_1 
&=\sum_{i=1}^N \exp -{2i\pi\over h}(e_i,\Gl) \qquad\qquad\qquad\quad
= \chi_1(M)\cr
	 \gamma^{(\Gl)}_2  
&=\sum_{1\le i<j\le N} \exp -{2i\pi\over h}
					((e_i+e_j),\Gl)
\quad\quad =\chi_2(M) \cr
		\vdots	\quad		&\qquad\quad \vdots
\qquad\qquad \vdots 
\cr  
 	\gamma^{(\Gl)}_{N-1} 
	&=\sum_{1\le i_1<\cdots i_{N-1}\le N}
	\exp -{2i\pi\over h}((e_{i_1}+\cdots + e_{i_{N-1}}),\Gl) \cr
   	&=\(\gamma^{(\Gl)}_1\)^* \qquad\qquad\qquad\qquad\qquad\quad\quad
	=\chi_{N-1}(M)\ .
	\cr }  }
%
We  call $h$ the ``Coxeter number" of the graph. Also we 
call these weights $\Gl$ ``exponents", and denote their 
set $\Exp$. 
Some of these exponents may occur with multiplicities. 
The class of graphs for which all the exponents belong to 
$\CP^{(h)}_{++}$  is referred to as
 class II, those for which they belong to $\CP^{(h)}_{+}$
(with some belonging to $\CP^{(h)}_{+} \setminus \CP^{(h)}_{++}$)
form the class I. 
\item{6)} We assume that the weight 0 for the class I, or $\Gr$ for the 
class II, is an exponent, with multiplicity 1. The corresponding 
eigenvalue reads 
\eqn\Ik{
\eqalign{ \Gc_p^{(0)} &={N \choose p} \qquad\ \hbox{ for \ class \ I} \cr
\hbox{ and \quad} \Gc_p^{(\Gr)} &={N \choose p}_{\!\!q} \qquad
\hbox{ for \ class \ II} \cr
\cr }}
%
in terms of a $q$-deformed binomial coefficient $q=e^{{i\pi\over h}}$, 
\eqn\Il{
{N \choose p}_{\!q} = {\sin {\pi N\over h}\cdots \sin {\pi (N-p+1)\over h}
\over \sin {\pi \over h}\sin {2 \pi \over h}\cdots \sin {p\pi \over h}}\ .}
%
\medskip
\nind {\it Remark 1}. By an abuse of language, what is called ``a graph" in the 
previous description is in fact the collection of $N-1$ graphs, 
with adjacency matrices $\CG_p$, 
connecting the given set of vertices $\CV$. Note that if the grading $\tau$ 
of each vertex is known, one may simply consider the unique graph
of adjacency matrix 
\eqn\Im{G= G_1+ \cdots + G_{N-1}\ .}
Each matrix $G_p$ is then identified as made of the matrix elements 
$G_{ab}$ of $G$ that satisfy $\Gt(b)-\Gt(a)=p\ \mod N$. 
\bigskip
\nind {\it Remark 2. } As mentioned in 2), it is natural to relax the 
assumption of integrality of the matrix elements $(G_p)_{ab}$ and to 
allow values of the form either $2\cos{\pi\over m_{ab}}$ ($m_{ab}\in \N$)
or ${N\choose p}_q$ (see \Il\ and sect. 3.4 below). 
By a small abuse of language, we still call $G$ the 
adjacency matrix of a graph, whose edges are decorated by some marking, 
{\it e.g.} $ \bullet \!\!{{m_{ab}}\over{\qquad}}\!\circ$
in the first case. This is a generalization 
of the distinction between simply and non simply laced 
Coxeter-Dynkin diagrams. Note that now 
the name ``(non) simply laced'' is inappropriate and should 
be replaced by ``(non) integrally laced'', as there are 
graphs whose adjacency matrix may have entries equal to 2 (see
examples below). 

\bigskip
\nind
%
\subsec{1.3 Examples} 
\nind 
1. Examples of class I for $N=2$ are provided by the affine Dynkin diagrams,
$\hat A_{2p-1}$, $\hat D_n$, $\hat E_6$, $\hat E_7$, $\hat E_8$. (The 
restriction to odd $2p-1$ in $\hat A$ comes from the assumption that the 
graph is bicolorable ($\IZ_2$ grading)). The conjugation $a\to  \ba$ is
trivial in these cases: $\ba \equiv a$. If the assumption of integrality
of the entries of the $G_p$ matrices is relaxed, one finds also
the $\hat B_n$, $\hat C_n$, $\hat F_4$ and $\hat G_2$ Dynkin diagrams. 
The $\hat A$, $\hat D$ and $\hat E$ cases are known to be in 
one-to-one correspondence with the finite subgroups of $SU(2)$ [MK].

The Coxeter number and exponents of the $\hat A, \hat D$ and $\hat E$ 
cases are as follows 
\halign{ # & # &  # \cr
\qquad \qquad
  $\hat A_{2n-1}$ &\qquad   $n$ &\qquad
 $0, 1, \cdots n-1, n, n-1, \cdots 1.$ \cr
\qquad \qquad
 $\hat D_{n+2}$ &\qquad $2n$ & \qquad $0, 2, \cdots 2n; n, n $   \cr
\qquad \qquad
 $\hat E_6$ &\qquad 6 &\qquad 0,2,2,3,4,4,6 \cr
\qquad \qquad
 $\hat E_7$ &\qquad 12 &\qquad 0,3,4,6,6,8,9,12 \cr
\qquad \qquad
 $\hat E_8$ &\qquad 30 &\qquad 0,6,10,12,15,18,20,24,30 \cr }

More generally, any finite subgroup $H$ of $SU(N)$ containing its 
center $\IZ_N$ yields a solution of class I. The set 
$\CV= \{R_a; a=1, \cdots , n\}$ is the 
set of irreducible representations of $H$. As $H$ is embedded into 
$SU(N)$, one also knows $N-1$ 
special representations $\CR_p$ of $H$, obtained by restriction of 
the $N-1$ fundamental representations of $SU(N)$ to $H$. Note that 
these representations are not necessarily irreducible. The matrix 
$G_p$ is then defined as providing the decomposition into irreducible
\rep s of the tensor product by $\CR_p$
\eqn\Ima{
\CR_p \otimes R_a  = \oplus_b (G_p)_{ab} R_b \ .}
The axioms 1)-6) are satisfied. The involution $a\mapsto \bar a$ 
is the conjugation of representations. The grading $\tau(a)$ 
is the ``$N$-ality" of the representation $R_a$
defined by 
\eqn\In{
\hbox{\rm if} \ \xi \in \IZ_N \subset H \quad R_a(\xi)= \xi^{\Gt(a)} \II }
and satisfies $\Gt(\bar a)=-\tau(a) \mod N$.
The matrices commute, as a consequence of the commutativity of the 
tensor product. The graph $\CG_1$ is connected: as the 
$N$-dimensional fundamental \rep\ is faithful, 
 all \rep s of $H$ appear in the decomposition 
 of  its repeated tensor powers [FH]. 
Properties 3) and 4) are simple consequences of the 
fact that $(G_p)_{ab}$ is the multiplicity
of the identity representation in the product $\CR_p\otimes R_a\otimes 
R_{\bar b}$. The eigenvalues of the matrices $G_p$ are the
fundamental characters of $SU(N)$, $\chi_p(M_\Ga)$, evaluated 
for a representative of the class $\Ga$ of $H$. As $\chi_p$ is
a class function in $SU(N)$, one may compute  $\chi_p(M_\Ga)$ in 
terms of the eigenvalues of $M_\Ga \sim \diag(\exp 
\Ge_j)_{j=1, \cdots, N}$,  $\sum_j \Ge_j=0$. These eigenvalues have the 
form $\Ge_j =-{2i\pi n_j\over |H|} $,  $n_j\in \IZ$, since 
any element of $H$ is at most of order $|H|$: $M_\Ga^{|H|}=\II$; 
the $n$'s are defined modulo $|H|$ and up to a permutation. 
One may identify 
\eqn\Io{
\eqalign{
 {n_i \over |H|}&= {(e_i,\Gl)\over h}
\qquad\hbox{for some $\Gl \in \CP$ and some $h \in \IN$}\cr
&= {1\over hN}\(\sum_{j=i}^{N-1} \Gl_j(N-j) -\sum_{j=1}^{i-1}j\Gl_j\) \cr}}
%
by taking $\Gl_i= h {(n_i-n_{i+1} )\over |H|} $. $h$ is chosen to be
the smallest common denominator of the fractions 
$ {(n_i-n_{i+1} )/ |H|}$, i.e. $h$ always divides $|H|$. 
The resulting $\Gl$ is not necessarily in $\CP_+^{(h)}$. 
One may however find an image $\Gm$ of $\Gl$ by the affine Weyl group of
$sl(N)$ such that $\Gm\in \CP_+^{(h)}$ and 
   $\gamma_p^{(\Gl)}=\gamma_p^{(\Gm)}$
(see below the proof of lemma 1.1).  This 
completes the proof that axiom 5 is indeed satisfied. Finally any 
subgroup of $SU(N)$ contains the class of the identity for which 
$\Gl=0$, with multiplicity 1.

\fig{\rmn The graph associated with the subgroup $\Sigma(1080)$ (a) 
and its truncated counterpart of class II (b).
The three colours --black, grey and white-- mark vertices with 
$\tau=0,1,2$ respectively.  The exponents of graph (b) are 
$(1,1),(3,3),(5,5)$ and their $\Gs$-orbit and twice $(4,4)$.}
{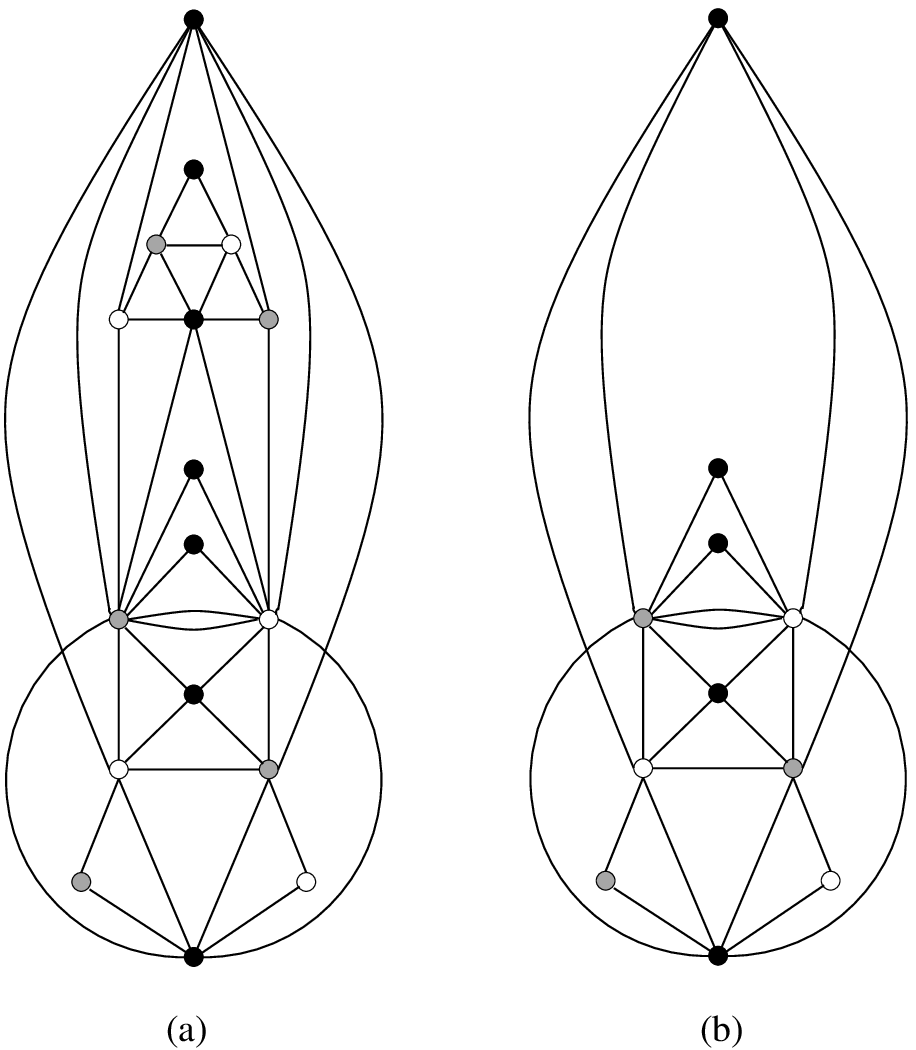}{75mm}\figlabel\subgr

As an example of a case with $N>2$, the graph associated with the subgroup
$\GS(1080)$ of $SU(3)$ (with the notations of [FFK])
is exhibited in Fig.\subgr(a). Its Coxeter number 
is $h=60$ and its $n=17$ 
exponents are $(0,0)$, $(15,15)$, $(30,30)$, $(6,6)$, $(12,12)$
and their transforms by $\Gs$, and $(10,10)$ with multiplicity 2. 
Drawings of more  graphs pertaining to $SU(3)$ may be found
in [DFZ1]. 
 
\bigskip

\medskip
\nind 2. Examples of class II 
for $N=2$ are provided by the ordinary simply laced Dynkin diagrams,
$ A_n$, $ D_n$, $ E_6$, $ E_7$, $ E_8$. 
 If the assumption of integrality
of the entries of the $G_p$ matrices is relaxed, one finds also
the $ B_n$, $ C_n$, $F_4$, $ G_2$, $H_3$, $H_4$ and $I_2(h)$ 
Coxeter-Dynkin diagrams, with the identifications $B_n=C_n$ 
(between their {\it symmetrized} Cartan matrices), and $G_2=I_2(6)$.
This in fact exhausts all the (bicolorable) graphs 
satisfying the axioms for $N=2$ [GHJ]. 
\smallskip
There exists another class of solutions known for all $N$, 
namely the fusion graphs of the affine algebra $\slh(N)$ at level
$k$. The vertices are the integrable weights described above, 
i.e. $\CV = \CP^{(h)}_{++}$, $h=k+N$. The definitions of the involution and 
of the grading are the same as in Example 1: the involution 
is $\Gl \mapsto \bar \Gl$ defined in \Ic\ and the grading may be
chosen as $\tau(\Gl)=\sum_{j=1}^{N-1} j(\Gl_j-1)\ \mod N$. 
The matrices $G_p$ are the Verlinde matrices, which describe the 
{\it fusion}, rather than the ordinary tensor product, by the 
$p$-th fundamental representation. Their
diagonalization is known, thanks to the Verlinde formula  [Ve], 
to be achieved by the unitary matrix $S_\Gl^{\ \Gm}$ 
of modular transformations of affine characters  [KP], and the
eigenvalues are the $\Gc_p^{(\Gl)}$, where $\Gl$ takes itself all the 
values in $\CP_{++}^{(h)}$. (This case enjoys self-duality properties, 
in agreement with the fact that 
both the vertex set $\CV$ and the set of exponents $\Exp$ 
are isomorphic to $\CP_{++}^{(h)}$.) In the case $N=2$, these 
fusion graphs reduce to the $A_{h-1}$ Dynkin diagram.
\smallskip
\fig{\rmn 
The first fusion graphs of $\slh(3)$. Same convention as in Fig.~\subgr}
{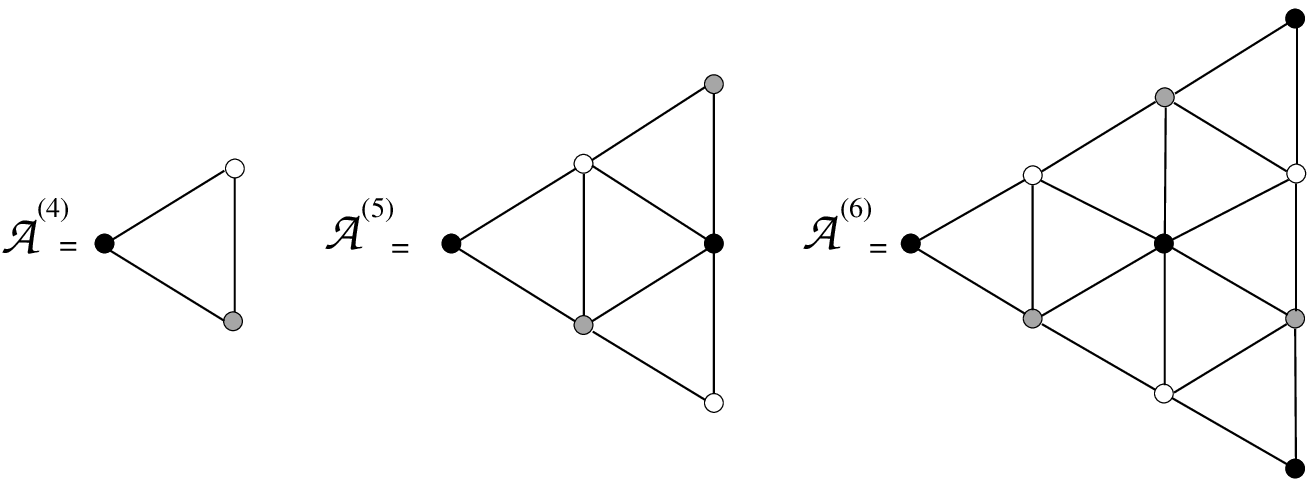}{100mm}\figlabel\fussutr

\fig{\rmn The first orbifolds of $\slh(3)$.
$\CD^{(6)}$ is the $\IZ_3$ orbifold of the graph $\CA^{(6)}$ of Fig. 
\fussutr. }
{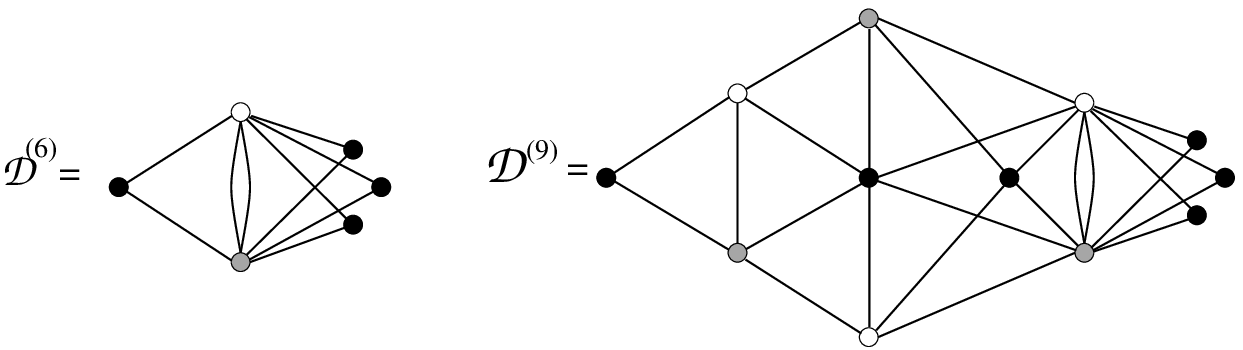}{100mm}\figlabel\orbsutr

More solutions are known, which generalize the two previous cases
of $ADE$ Dynkin diagrams and of fusion graphs. Infinite series
--which are the analogues of the $D$ Dynkin diagrams-- have been 
obtained by orbifolding the fusion graphs [Ko], see Fig \orbsutr\foot
{\rmn It is amusing to notice that the graph $\CD^{(6)}$
has already been encountered in 
another systematic study of ``extended affine root systems" [S]: 
this is the graph denoted there $D_4^{(1,1)}$. }. 
Several others have been 
obtained, originally by empirical methods [DFZ1], starting from
graphs of finite subgroups of $SU(N)$ (cf. Example 1) and 
removing some vertices and edges in order to attain the spectral 
properties of axioms 5-6. For example, a graph related to the 
previously subgroup $\Sigma(1080)$ is presented in Fig \subgr(b). It has 
$h=12$.  Recently, more systematic methods
have been set up, which make use of some data of rational conformal 
theory to construct  graphs [PZ]. There is, however, a shortage 
of general results on the class of graphs satisfying the previous
axioms. 

\bigskip\penalty -5000
\nind 
{\it Problems}: \penalty 10000
 \nind 1. Classify all the solutions to the axioms 1-6. 
\nind This is presumably too difficult. More modest questions are:
\nind 2. Are all  solutions of class I obtained from some 
finite subgroup of $SU(N)$? 
\nind 3. Are all solutions of class II obtained by truncation of some
solution of class I? 
\nind It appears that the answers to 2 and 3 cannot be both ``yes". 
In [DFZ1], solutions of class II were found for $N=3$ which do not seem 
to emanate from a known subgroup of $SU(3)$. 
%
%
%
%
%
\bigskip
\subsec{1.4 Simple consequences of the axioms}  %
\nind {\bf Lemma 1.1 : }{ \sl The set of $\Gc_p^{(\Gl)}$, 
for all $p=1, \cdots, N-1$, 
characterizes the weight $\Gl$ in $\CP_{+}^{(h)}$: 
\eqn\Ip{
{\sl if\ for\ all}\quad p=1,\cdots, N-1 \quad \Gc_p^{(\Gl)}=
\Gc_p^{(\Gm)}\  ,\quad {\sl then}\quad \Gl=\Gm\ .}
 }
{\it Proof} : The $\Gc_p^{(\Gl)}$, $p=1, \cdots, N$ (with $\Gc_N^{(\Gl)}=1$), 
are the elementary symmetric functions of the $ \ve_i(\Gl)=\exp-{2i\pi\over h}
(e_i,\Gl)$, $i=1, \cdots,N$. Thus $\Gc_p^{(\Gl)}=\Gc_p^{(\Gm)}$ for all 
$p=1, \cdots , N$ implies that $\ve_i(\Gl)$ is a permutation of the 
$\ve_i(\Gm)$. Since  the Weyl group of $sl(N)$ acts on the $e_i$ by 
permutation, this means that there is an element $w$ of the (finite) Weyl 
group such that 
\eqn\Iq{
\ve_i(\Gm)  
=  e^{-{2i\pi\over h}(e_i,\Gm)} 
= e^{-{2i\pi\over h}(we_i,\Gl)} 
= e^{-{2i\pi\over h}(e_i,w^{-1} \Gl)} \ , 
}
i.e. for all $i$, 
$(e_i,\Gm)=(e_i,w^{-1}\Gl)+n_i h $, ($n_i\in \IZ$, $\sum n_i=0$). 
Thus there is 
an element $\tilde w$ of the {\it affine} Weyl group such 
that $ \Gm=\tilde w \Gl$. Since $\CP_+^{(h)}$ is a fundamental domain 
of the weight lattice under the action of the affine Weyl group, 
we conclude that if both $\Gl$ and $\Gm$ belong to that domain, they are 
identical. q.e.d. By the same token, we learn that 
any weight $\Gl \in \CP$ has an image $\Gm$ by the affine 
Weyl group that belongs to $\CP_+^{(h)}$ such that 
$\Gc_p^{(\Gl)}= \Gc_p^{(\Gm)}$ for all $p$.

Now consider the set $\Exp$. One has the following 
\smallskip\nind 
{\bf Proposition 1.2:} {\sl The set $\Exp$ is invariant under the action 
of $\Gs$ and of $\CC$. In other words, the set $\Exp$ contains full orbits 
of $\Gs$ and of $\CC$.  Moreover the eigenvectors may be chosen so as to satisfy}
\eqna\Ir
$$
\eqalignno{
\psi^{(\bar\Gl)}_a &=\psi^{(\Gl)}_{\bar a} = \(\psi^{(\Gl)}_a\)^*  & \Ir a 
\cr
\psi^{(\Gs(\Gl))}_a&= e^{{2i\pi\over N}\tau(a)} \psi^{(\Gl)}_a\ .
& \Ir b\cr}
$$
This follows from the fact that $(e_i,\Gs(\Gl))=(e_{i-1},\Gl) -{h\over N}$
for $i=2,\cdots, N$ and $(e_1,\Gs(\Gl))=(e_{N},\Gl) +{N-1\over N}h$, whence 
\eqn\Ira{\eqalign{\ve_i(\Gs(\Gl)) &=e^{{2i\pi\over N}} \ve_{i-1}(\Gl)\cr
\ve_i(\Gl)^* &= \ve_{N+1-i}(\bar \Gl) \cr }}
and thus
\eqn\Is{
\eqalign{ \Gc_p^{(\Gs(\Gl))}&= e^{2i\pi{p\over N}} \Gc_p^{(\Gl)}\cr
\hbox{ and}\quad \Gc_p^{(\bar\Gl)}&= \(\Gc_p^{(\Gl)}\)^* \ ;\cr}}
%
then from axiom 3), if $\psi_a^{(\Gl)}$ is an eigenvector for the 
eigenvalue $\Gc_p^{(\Gl)}$ of $G_p$, all $p$,  so is $\tilde\psi_a^{(\Gl)}:=
e^{2i\pi{\tau(a)\over N}} \psi_a^{(\Gl)}$ for the eigenvalue 
$\Gc_p^{(\Gs(\Gl))}$. By lemma 1.1, this suffices to qualify $\Gs(\Gl)$ as
an exponent. Likewise, the reality of the characteristic polynomials 
implies that $\Gc_p^{(\Gl)}$ and  $\Gc_p^{(\Gl)*}$ are 
simultaneously eigenvalues, and thus $\Gl$ and $\bar\Gl$ are 
simultaneously exponents. 

\smallskip\nind 
{\bf Corollary 1.3: } {\sl The sum of exponents is}
\eqn\Isa{ \sum_{\Expexp} \Gl = {h \, n\over N}  \Gr }
where we recall that $n$ is the number of vertices and $\Gr$ is the 
sum of all the fundamental weights. 
\smallskip
\nind {\it Proof} : 
Compute $(1+\Gs+\cdots +\Gs^{N-1}) \sum \Gl$ in two different 
ways. On the one hand,  as $\Exp$ is invariant under $\Gs$, the sum equals
$N\sum_{\Expexp} \Gl$. 
On the other,  for each $\Gl$, $(1+\Gs+\cdots +\Gs^{N-1})\Gl=h\Gr$, 
thus the previous sum also equals $h n \Gr$.  

\smallskip\nind
{\bf Corollary of  corollary 1.3}: $N$ divides $h n$.
\medskip
For later use, I introduce  an explicit  parametrization 
of the matrices $G_p$. 
I assume that the vertices of $\CV$ have been ordered according to
increasing $\tau$: first the vertices with $\tau=0$, then $\tau=1$, etc.
Then the matrices $G_p$ are $N\times N$  block--matrices of the  form
\eqn\Isb{
G_p =\pmatrix{0  & \cdots & 0 &A_{1\,p+1}& \cdots & 0         \cr
            \vdots& \ddots      &  &        & \ddots & \vdots    \cr
		0 & \cdots      & 0 &\cdots   &   0    &A_{N-p\, N}\cr
     A_{N-p+1\, 1}& 0      &  & 0        &\cdots  & 0         \cr   
            \vdots& \ddots & & 0        &\ddots  & \vdots    \cr
    		0 & \cdots & A_{N\, p} & 0&        &  0        \cr   }\ ,
}
with the matrices $A_{ij}$ satisfying $A_{ij}^t=A_{ji}$
as a consequence of \Ig. 
(The matrices $A_{ij}$ are of course subject to further constraints
expressing the commutation of the matrices $G$, etc). 
Later, we shall also encounter the matrix%
\foot{\rmn Here and in the following, by a small abuse of notations,
$\un$ denotes a unit matrix, whose dimension is fixed by the context.}
\eqn\It
{T=\pmatrix{\un & A_{12}& A_{13}& \cdots &  A_{1N}\cr
		  0     & \un    & A_{23}& \cdots &  A_{2N}\cr
		       & 0      & \ddots &       &    \cr
		       &        &   0   & \un    & A_{N-1\, N}\cr
		0      &      0 & \cdots& 0      & \un \cr}\ .}
It may be written as a product of upper triangular matrices in the 
two following ways:
\eqn\Iu
{\eqalign{T&= \pmatrix{\un & & & & \cr
                                &\un & &0 & \cr 
				& & \ddots & &\cr
				&0 & & \un & A_{N-1\, N}\cr
				 & & & & \un \cr}
\cdots \pmatrix{\un & A_{12}& A_{13}& \cdots &  A_{1N}\cr
                    & \un &0 & &0 \cr
		    & & \ddots &0 & \cr
		    &0 & & \un & 0\cr
		    & & & & \un  \cr } \cr
&= \pmatrix{\un & & & &  A_{1N}\cr
             & \un & &0 & A_{2N}\cr
	     &  &\ddots & &\vdots \cr
	     &0 & & \un & A_{N-1\, N}\cr
	     & & & & \un \cr}\cdots 
\pmatrix{\un & A_{12}&0 & \cdots & 0\cr
            & \un & & 0 & \cr
	    & & \ddots & & \cr
	    &0 & & \un & \cr
	    & & & & \un \cr    }\ , \cr}}
%
which allows to write its inverse and its transpose as
%
%
\eqn\Iv
{T^{-1}= \pmatrix{\un & -A_{12}& -A_{13}& \cdots & - A_{1N}\cr
                    & \un &0 & &0 \cr
		    & & \ddots &0 & \cr
		    &0 & & \un & \cr
		    & & & & \un  \cr } \cdots
		    \pmatrix{\un & & & & \cr
                                &\un & &0 & \cr 
				& & \ddots & &\cr
				&0 & & \un & -A_{N-1\, N}\cr
				 & & & & \un \cr} }
%
and
%
\eqn\Iw
{ T^t= \pmatrix{\un & &0 & \cdots & 0\cr
     	    A_{21} & \un & & 0 & \cr
	    & & \ddots & & \cr
	    & 0& & \un & \cr
	    & & & & \un \cr    }
\cdots
\pmatrix{\un & & & & \cr
             & \un & &0 & \cr
	     &  &\ddots & & \cr
	     & 0& & \un & \cr
	     A_{N1} & A_{N2} &\cdots &A_{N\,N-1} & \un \cr} \ .}


\penalty -5000

\AHead{2. Reflection Groups and Root Systems} %
\secno=2
\penalty 10000
\subsec{2.1 Roots, bilinear form and reflection group.} 
\nind
Given a graph of the previous type, we associate with it a 
vector space $V$ over $\IR$, with a basis $\{\Ga_a\}$ 
labelled by the vertices $a$ of the set $\CV$.
A bilinear form is defined on the vectors of that basis by
\eqn\IIa{
g_{ab} = \bra \Ga_a, \Ga_b\ket = 2 \Gd_{ab} + G_{ab}}
%
in terms of the matrix $G =\sum_{p=1}^{N-1} G_p$. 
\foot
{\rmn Note that the sign differs from that used in [Z] if $N$ is even, 
and in particular for $N=2$, 
from that commonly used for the finite reflection groups for
which $g$, the Cartan matrix, is rather written as $2 \Gd_{ab}-G_{ab}$. 
One passes from one sign to the other by the change 
$\Ga_a \to (-1)^{(N-1)\Gt(a)} \Ga_a $. This new convention has the merit
of sparing a lot of irrelevant signs and of unifying the cases of 
$N$ even and odd. One should however recall this unorthodox choice of sign 
when examining the case $N=2$. 

Note also that according to the usual
conventions, all the non negative scalar products of roots in \IIa\
should be represented by {\nineit broken} lines in the generalized Dynkin 
diagrams.  For the ease of drawing, we use solid lines.  }
In terms of the matrix $T$ introduced in \It, one has simply
%
\eqn\IIb{g= T + T^t \ .}
%
\nind
The vectors $\Ga_a$ will be called {\it roots}. 
For each root $\Ga_a$, one considers the reflection
in the hyperplane through the origin orthogonal to $\Ga_a$
%
\eqn\IIc{S_a: x \in V \mapsto x'= S_a x= x - \bra \Ga_a,x\ket \Ga_a \ .}
%
In components, if $x= \sum_c x_c \Ga_c$, then
\eqn\IId{\eqalign{
x'_a&= -x_a -\sum_{c\ne a} g_{ac} x_c = -x_a -\sum_{c\ne a} G_{ac} x_c\cr
x'_b&= x_b \qquad \hbox{ if}\ \ b\ne a \ .}}
%
One then considers the group $\GC$ generated by the $S_a$, $a=1, \cdots,n$. 

Note that $S_a^2= \II$; if $G_{ab}=1$, then $(S_aS_b)^3=\II$ and more
generally if $G_{ab}=2 \cos {\pi p\over m}$, then $(S_a S_b)^m=\II$. 
The generators $S_a$ of the group may however satisfy additional relations 
[Z]. This would make the denomination ``Coxeter group" improper, although 
as we shall see such groups share several properties with Coxeter groups.
 
\bigskip

The signature of the metric $g$ is of central importance. In particular 
it is well known that the group $\GC$ is of finite order if and only if the 
metric is positive definite. 
In the present case, one knows the eigenvalues of $g$ 
%
\eqn\IIe{ g^{(\Gl)}= \sum_{p=0}^N \Gc_p^{(\Gl)}}
%
with the convention that $\Gc_0^{(\Gl)}= \Gc_N^{(\Gl)}=1$. 
Using \Ij, it takes the multiplicative form

%
\eqn\IIf{
g^{(\Gl)}= \prod_{i=1}^N \( 1+e^{-{2i\pi\over h} (e_i,\Gl)} \)\ .}
%
Then one has the following
\smallskip
\nind {\bf Proposition 2.1 :} {\sl The form $\bra \ , \ \ket$ is positive 
definite if and only if the graph is of class II and }
\item{} $N=2 \qquad \forall h\ge 3$
\item{} $N=3 \qquad h=4,5$
\item{} $N\ge 4 \qquad h=N+1$ . 
\smallskip
\nind The principle of the proof is to use the fact that the 
weight $0$ in class I, or $\Gr$ in class II, is known to belong to 
the set of exponents, together with its $\Gs$-orbit. Thus for class I,
\eqn\IIg{
\Gc_p^{(\Gs^\ell(0))} = e^{2i\pi{p\ell\over N}} {N \choose p}}
and 
\eqn\IIh{g^{(\Gs^{\ell}(0))} =(1+e^{2i\pi{\ell\over N}})^N= 
(-1)^\ell \(2\cos{\pi\ell\over N}\)^N }
%
For $\ell=1$, this vanishes  for $N=2$, and is negative for $N>2$. 
A similar discussion may be carried out for the class II, looking 
at the orbit of $\Gr$. The details have been provided in [Z]. 

In that same reference [Z], the graphs which satisfy the conditions 
of Proposition 2 have been spelled out. For $N=2$, one recovers 
of course the $ADE$ Dynkin diagrams, together with the other 
Coxeter-Dynkin diagrams if the integrality of the matrix elements 
is relaxed. For $h=N+1$, the only graph is the fusion graph of 
$\slh(N)$ at level $k=1$. At that level, the integrable weights
are $\Gr$ and the fundamental weights shifted by $\Gr$, $\GL'_i:=\GL_i+\Gr$, 
$i=0, \cdots, N-1$, $\GL_0:=0$. The fusion ring of $\slh(N)_1$ 
is isomorphic to the cyclic group $\IZ_N$,  with $G_p$ acting as a 
cyclic permutation of these weights, 
$\GL'_i\mapsto \GL'_{i+p\ \modmod   N}$, and the graph of $G_p$ 
is thus a star-polygon. For $N=3$ and $h=5$, two 
solutions are known. The first has $n=3$ vertices, 
and all off diagonal entries of $G$  are $G_{ab}=2 \cos {\pi\over 5}$. 
The second has 6 vertices, it is the fusion graph of $\slh(3)$ at level 2. 
It is most likely that those are the only solutions [Z]. 

\medskip
The form $\bra \ , \ \ket$ is thus generically indefinite and the 
group   of infinite order. In general, 
it may be proved that the number of 
negative eigenvalues of the bilinear form  $g$ is even [Z].  
Moreover, for graphs of class II, the number of zeros is also even.
(See also a conjecture on the signature at the end of next
subsection.)
A few identifications or isomorphisms 
 of such infinite  groups have been obtained 
in reference [Z], to which we refer the reader. 	
See also some results in sect. 4 below. 

%
\bigskip%
\subsec{2.2 Coxeter element} %
\nind The properties of the Coxeter element, product of all the reflections 
relative to simple roots  of finite Coxeter groups, are well known~[C]. 
There is a weak analogue in 
the more general class of reflection  groups which we are considering here. 
One can make use of the grading of vertices to define
the following product of reflections
%
\eqn\IIi{
R= \prod_{\tau(a)=0} S_a  \prod_{\tau(b)=1} S_b  
\cdots\prod_{\tau(f)=N-1} S_f \ . }
%
\smallskip
\nind {\bf Proposition 2.2 : } {\sl 
The element $R$ is independent of the order  of the $S$ within each block; 
it is conjugate in the linear group $GL(n)$ to the product $-T^{-1} T^t$ of the matrices
defined at the end of sect. 1.4, and its spectrum is of the form}
%
\eqn\IIj{ 
(-1)^N \exp N {-2i\pi\over h}(e_j,\Gl)\qquad \Gl\in \Exp\, , \quad 
\forall j\ \hbox{\rm fixed: }\  1\le j \le N \ . }
%
\smallskip
\nind (Indeed this set of numbers is independent of $j=1,\cdots, N$. ) 
The proof is a simple extension of the original proof by Coxeter [C]. It 
has been given in [Z] (up to changes of signs due to the 
changes of conventions in the metric), and it will be only sketched here.
On the one hand, one proves that the $i$-th term in the product \IIi\  
equals
%
\eqn\IIk{
\eqalign{\prod_{\tau(b)=i-1} S_b  
&=\pmatrix{\un & & & & \cr
	  &  \ddots & & & \cr
	  & &  \un & -A_{i\,i+1} &\cdots & -A_{i\,N} \cr
	&0 & &\ddots & \cr
	& & & & \un \cr}
\pmatrix{\un & && & \cr
	   & \ddots & & & \cr
	-A_{i1}& \cdots& -\un & & \cr
	 & 0 & &\ddots& \cr
	 & & & & \un \cr}  \cr
& =: B_i C_i \cr } }
%
and  that $B_i$ commutes with $C_j$, $1\le j<i\le N$. Since
 $B_1 \cdots B_N = T^{-1}$  and  $ C_1 \cdots C_N = - T^t$, 
the product $R$ is equal to the product $- T^{-1} T^t$. 
On the other hand,  the polynomial with roots 
$ -\ve_j^{(\Gl)}=-\exp- {2i\pi\over h}(e_j,\Gl)$
reads 
\eqn\IIl{{\eqalign
{\GD(z)&=\prod_{j=1}^N\prod _{\Gl\in \Expexp} \(z+\ve_j^{(\Gl)}\) \cr
&=
\det\( z^N \un +z^{N-1} G_1 + z^{N-2} G_2 +\cdots + zG_{N-1}+ \un\)\cr
&= \det \pmatrix{
(z^N+1)\un & z^{N-1} A_{12}& z^{N-2} A_{13}& \cdots & z A_{1N}\cr
  z A_{21}    & (z^N+1)\un & z^{N-1}A_{23}&     & z^2 A_{2N} \cr
\vdots & & &\ddots & \cr
z^{N-1}A_{N\, 1} & z^{N-2}A_{N\, 2} &\cdots & &  (z^N+1)\un \cr }  \cr
 &= \det(z^N T +T^t)= \det (z^N \un  - (-T^{-1}T^t))\ ,\cr }}}
using \Ij\ and \It. This proves that the eigenvalues of 
$R=- T^{-1} T^t$ are the $(-\ve_j(\Gl))^N$, which are indeed 
independent of $j$. 
%

\smallskip
 It is an interesting question to know if this Coxeter element 
is of finite or infinite order. Even though its eigenvalues 
are roots of unity, $R$ may be non-diagonalizable
and thus of infinite order. One has the following
\smallskip

\nind {\bf Proposition 2.3 : } {\sl For any graph of class I, 
the Coxeter element $R$ is of infinite order. }

\noindent The obstruction to diagonalizability of $R$ 
comes from its eigenvalue
$(-1)^N$, which is associated by Proposition 2.2 with the exponent $0$ and 
its $\Gs$ orbit, and  which has thus a multiplicity at least equal to $N$. 
The proof consists in finding a subspace of $V$ in which $R$
takes a triangular form with this eigenvalue on the diagonal. Let  
\eqn\IIm{\Gb_i=\sum_{a: \Gt(a)=i} \psi_a^{(0)} \Ga_a\ , \qquad\qquad i=0,\cdots, N-1}
and 
\eqn\IIn{\Gd_k =\sum_{i=k}^{N-1} (-1)^{i-k} {i\choose k} \Gb_i \ , \qquad\qquad 
k=0, \cdots, N-1\ ,}
for $i,k=0, \cdots, N-1$. Then 
denote $R_j:=\prod_{\Gt(a)=j} S_a$. 
One computes easily, using the eigenvalues $\Gc^{(0)}_p$ of \Ik, 
\eqn\IIo{\eqalign{ R_j \Gb_i &= \Gb_i -{N\choose |i-j|}\Gb_j-\Gd_{ij}\Gb_i\cr
R_j \Gd_k &= \Gd_k -(-1)^k\Big[ (-1)^i {j\choose k}+
 \sum_{i=0}^{N-1} (-1)^i {i\choose k} {N\choose |i-j|}   \Big]\Gb_j \cr}\ . }
Then it is tedious but straightforward to establish by induction that
\eqn\IIp{\eqalign{
R_j R_{j-1}\cdots R_0 \Gd_k&=
\Gd_k  -\sum_{i=0}^{j-k}(-1)^i
{i+k\choose k}\Gb_{i+k}  \cr
& \qquad
+(-1)^{N-k}\sum_{i=0}^j \( {N+i\choose k}-\sum_{\ell=j+1}^k 
{i\choose \ell} {N\choose k-\ell} \)\Gb_i  \cr}}
in which the sum $\sum_{\ell=j+1}^k$ (resp $\sum_{i=0}^{j-k}$)
 reduces to naught for $j \ge k$ (resp $k\ge j$).
Taking $j=N-1$ in \IIp, it follows that 
\eqn\IIq{\eqalign{R^{-1} \Gd_k = R_{N-1}\cdots R_0 \Gd_k
&=  (-1)^{N-k}\sum_{i=0}^{N-1} {N+i\choose k}
\Gb_i \cr 
&= (-1)^N \(  \Gd_k +\sum_{i=1}^k (-1)^i {N \choose i} \Gd_{k-i}\) \ .
 \cr }}
This is the triangular form referred to above and this clearly 
contradicts the possibility that $R$ be of finite order. 
Note that the previous proof is in essence the generalization to 
$N$-colourable graphs of the proof by Berman, Lee and Moody for 
bi-partite graphs [BLM]. 

In contrast with this case of the class I, it appears that 

\nind {\bf Proposition 2.4 : } {\sl For any graph of class II, 
the Coxeter element is of  finite order. }

This has been proved  by V. Petkova using a different 
approach that extends the discussion of Kostant [K] and that
also yields the result of Proposition 2.3 [Pe].

If the  Coxeter element $R$ is of finite order, this order is equal 
{\it at most} to $h$ if $Nh$ is even and to  $2h$ if $Nh$ is odd. 
The order may be smaller, e.g. for the graphs $\CD^{(6)}$ and 
$\CD^{(9)}$ of Fig. \orbsutr, 
 it is 2 resp. 6, while the value of $h$ is given by the superscript. 
If $m$ is a common divisor of $N$ and $h$ such that $\forall \Gl\in \Exp$, 
$\tau(\Gl)=0 \mod m$, then 
the order of $R$ is $h/m$ (resp $2h/m$) if $N(N-1+h)/m$ is even (resp odd)
[Pe]. 
%

The ordering chosen in \IIi\ is the only natural  one for most cases. 
In the case of fusion graphs, however, it is possible to use a 
lexicographic ordering of the vertices, namely  of
the integrable weights $\Gl$ at some 
level $k$ of $\slh(N)_k$. Thus for $\Gl,\Gm\in \CP^{(h)}_{++}$ with 
the notations of equ \Ia, one writes $\Gl \prec \Gm$ if
$\Gl_{N-1}=\Gm_{N-1}, \cdots \Gl_p=\Gm_p$ and $\Gl_{p-1} <\Gm_{p-1}$.
One verifies for the lowest levels, and it is quite likely to be true in
general, that the product of the $S$'s according to this ordering 
defines another Coxeter element equivalent to \IIi. 
This definition of the Coxeter element 
 is more natural from the point of view of the monodromy 
of the associated singularity (see below sect.4).

The eigenvalues of the Coxeter element as given by Prop. 2.2 
seem to specify the signature of the metric, in the case
of class II graphs. For each exponent of $\Gl$ define the real numbers
%
\eqn\IIr{\eqalign{
q_\Gl^{\hbox{\sevenrm (R)}} &= -{N\over h}(e_N,\Gl) -{N-1\over 2} \cr
&=  {1\over h}\sum j(\Gl_j-1) 
+{(N-h)(N-1)\over 2h}\ . \cr }}
Note that the eigenvalues of 
$-R$ are $\exp 2i\pi  q_\Gl^{\hbox{\sevenrm (R)}}$, for $\Gl\in \Exp$.
Then 
\medskip\noindent
{\bf Conjecture 2.5} : 
{\sl The signature of the metric $g$ for class II graphs
is $(r\, +, s\, -, t\, 0)$  where $r$ is the number of 
$q_\Gl^{\hbox{\sevenrm (R)}}$ which fall in an interval
$]2p-\oh, 2p+\oh[$, $p\in \IZ$, $s$ the number of those in an interval
$]2p+\oh, 2p+{3\over 2}[$, and $t=n-r-s$ the number of those
which are half-integers.}
\smallskip\nind
Alternatively, the signature is obtained by looking at the signs
of $\hbox{Re}\,\,\exp i\pi{q_\Gl^{(R)}}$.
This is suggested by the work of 
Steenbrink [St] on the intersection form of quasi-homogeneous 
singularities and by the one of Cecotti and Vafa  [CV] on 
$\CN=2$ supersymmetric field theories and topological field theories. 
In the latter, $q_\Gl^{\hbox{\sevenrm (R)}}$ are the 
	$U(1)$ charges of the Ramond ground states. 
Cecotti and Vafa study the monodromy operator of a linear 
system related to the $t t^*$ equations. The previous
conjecture follows from their work if one identifies
their monodromy operator with the present Coxeter element, 
which seems highly plausible. 
The last part of the conjecture is easy to prove: 
using once again eq. \Ira\ and the product formula \IIf\ for
$g^{(\Gl)}$,  
one proves that on each $\Gs$-orbit of an exponent $\Gl$, 
there are as many vanishing $g^{(\Gl')}$ as there are 
 $\ve_N(\Gl'')^N$ equal to $(-1)^N$, {\it i.e.}
 half-integral $q_{\Gl''}^{(R)}$.

 Conjecture 2.5  has been tested on all known 
examples.


%
%
\AHead{3. Pasquier algebras, subalgebras and folding.} %
%
\subsec{3.1 Pasquier algebras and (quasi) C-algebras} %
\nind 
For any graph of the previous type, the eigenvector $\psi^{(0)}$
for the class I, or $\psi^{(\Gr)}$ for class II, is the eigenvector
of $G_1$ of largest eigenvalue, i.e. the so-called Perron-Frobenius
eigenvector. As the graph is connected 
(cf axiom 2), all the components of the 
Perron-Frobenius eigenvector are non-vanishing and may be 
chosen real positive. We shall denote $\o$ this special  exponent 
$0$ (class I) resp $\Gr$ (class II).

Now consider the numbers [P]
%
\eqn\IIIa{
M_{\Gl\Gm}^{\ \ \Gn}=\sum_a {\psi^{(\Gm)}_a \psi^{(\Gn)}_a \psi^{(\Gn *)}_a
\over \psi^{(\o)}_a }\ . }
%
Since the $\psi$ are orthonormal, $M_{\o \Gm}^{\ \ \Gn}=\Gd_{\Gm\Gn}$. 
Since they satisfy 
$\psi_a^{(\Gl)*}=\psi^{(\bar \Gl)}_a=\psi^{(\Gl)}_{\bar a}$, 
the $ M_{\Gl\Gm}^{\ \ \Gn}$ are real,  the involution in $\Exp$: 
$\Gl \mapsto \bar \Gl$ is an automorphism 
$ M_{\Gl\Gm}^{\ \ \Gn}= M_{\bar\Gl\bar\Gm}^{\ \ \bar\Gn}$, 
and 
\eqn\IIIb{
\eqalign{
 M_{\Gl\Gm}^{\ \ \o}&=\sum_a  \psi^{(\Gl)}_a \psi^{(\Gm)}_a \cr
&= \sum_a  \psi^{(\Gl)}_a \psi^{(\bar\Gm)*}_a = \Gd_{\Gl\bar \Gm}\ .\cr}}
%
The algebra over $\IC$ defined by
these structure constants  $ M_{\Gl\Gm}^{\ \ \Gn}$ is a commutative 
and associative (as is readily verified) algebra, called the 
Pasquier algebra relative to the graph. 
It is  a quasi-C-algebra in the following sense: 

\medskip
\nind {\bf Definition : }
An algebra $\CA$ over $\IC$ with a given basis $x_1, \cdots, x_n$, 
is a {\it quasi C-algebra} if it satisfies the following axioms:
\item{i)} it is a commutative and associative algebra with {\it real}
structure
constants $p_{\Ga\Gb}^{\ \ \Gc}$, i.e. $x_\Ga. x_\Gb =
\sum_\Gc p_{\Ga\Gb}^{\ \ \Gc} x_\Gc$;
\item{ii)} it has an identity element, denoted $x_1$, i.e. 
$p_{1\Ga}^{\ \ \Gb}=\Gd_{\Ga\Gb}$;
\item{iii)} there is an involution on the generators $x_\Ga \mapsto
x_{\bar \Ga}$ which is an automorphism of the algebra, i.e.
$p_{\Ga\Gb}^{\ \ \Gc}=p_{\bar \Ga\bar \Gb}^{\ \ \bar \Gc}$;
\item{iv)} $p_{\Ga\Gb}^{\ \ 1}= k_\Ga \Gd_{\Ga\bar \Gb}$, with $k_\Ga$ a real 
positive number: $k_\Ga>0$.

\smallskip
\nind {\bf Definition : }[BI]\ A {\it C-algebra} is a quasi C-algebra 
which satisfies the additional property
\item{v)} the $k_\Ga$ form a one-dimensional representation of the algebra. 

\smallskip
The theory developed by Bannai and Ito may be summarized as follows. 
The axioms i-iv) imply that a quasi-C-algebra is semi-simple. Let 
$e_i$, $i=1,  \cdots, n$ denote the idempotents $e_i.e_j= \Gd_{ij} e_i$
and decompose 
\eqn\IIIba{x_\Ga=\sum_i p_\Ga(i) e_i \ . }
If axiom v) is also satisfied, one may choose $p_\Ga(1)=k_\Ga$. Then
one may introduce a second 
multiplication on $\CA$ for  which the original generators act as
idempotents $x_\Ga\circ x_\Gb= \Gd_{ab} x_a $. By inversion  of \IIIba\ 
this yields a non trivial   $e_i\circ e_j= q_{ij}^{\ \ k} e_k$. 
$\CA$ is also a C-algebra for this second multiplication, and 
there are thus {\it two} dual C-algebra structures defined on 
the set $\CA$. 

One has a nice illustration of this in the present context of graphs.
It is clear that the axioms of a quasi C-algebra are satisfied 
by the $ M_{\Gl\Gm}^{\ \ \Gn}$. On the other hand, it is easy to verify
that one knows $n$ 1-dimensional representations of the algebra, 
namely 
%
\eqn\IIIc{ 
\sum_\Gn M_{\Gl\Gm}^{\ \ \Gn} {\psi^{(\Gn)}_a\over \psi^{(\o)}_a}=
{\psi^{(\Gl)}_a\over \psi^{(\o)}_a}{\psi^{(\Gm)}_a\over \psi^{(\o)}_a}
\ .}
%
Thus if there exists one vertex, call it 1, such that $\psi^{(\Gl)}_1$
are non vanishing  
for all $\Gl\in \Exp$, then 
at the possible price of an overall change of sign one may assume 
them positive and 
%
\eqn\IIId{
\sqrt{k_\Gl}:={\psi^{(\Gl)}_1\over \psi^{(\o)}_1} }
%
and $p_{\Gl\Gm}^{\ \ \Gn}= \sqrt{{k_\Gl k_\Gm\over k_\Gn}}
M_{\Gl\Gm}^{\ \ \Gn}$ 
satisfy all the conditions of a C-algebra, including v). The quantities 
${ \psi^{(\Gl)}_a\over \psi^{(\Gl)}_1}$ are the one-dimensional 
\rep s of a dual algebra with structure constants 
%
\eqn\IIIe{
N_{ab}^{\ \ c}= \sum_{\Gl\in \Expexp} {\psi^{(\Gl)}_a\psi^{(\Gl)}_b\psi^{(\Gl)
*}_c \over \psi^{(\Gl)}_1} \ .}
The algebra $\CA^*$ with structure constants $q_{ab}^{\ \ c}= \sqrt{k^*_a
k^*_b/ k^*_c} N_{ab}^{\ \ c}$, with $\sqrt{k^*_a}= 
\psi^{(\o)}_a/\psi^{(\o)}_1$, 
 is a C-algebra, and $\CA$ and $\CA^*$ form a pair of dual C-algebras. 
The most favorable case is when the 
structure constants of both algebras are non negative (see below).
\bigskip
In fact the three possibilities occur in our graphs        :
\item{1.} For a given choice of the $\psi$, 
non existence of a vertex $a$ such that $\forall \Gl, \
\psi^{(\Gl)}_a >0$. This is for example the case of the 
$\CD^{(9)}$ diagram of Fig. \orbsutr\ if we take for the exponent $\Gl=(3,3)$
eigenvectors with components on the three rightmost 
vertices equal respectively to $(0,0,0)$, $(1,\Go,\Go^2)/\sqrt{3}$ and
$(1,\Go^2,\Go)/\sqrt{3}$, with $\Go=\exp {2i\pi \over 3}$. 
\item{2.} Existence of such a vertex, but the resulting $M$
and $N$ structure constants are of either sign. This is for 
example the case of the $D_{2\ell +1}$ or $E_7$ Dynkin diagrams.
This is also the case of the graph of Fig. \subgr(b). 
\item{3.} Existence of such a vertex, and the $M$ and $N$ (and $p$ and $q$)
structure constants are non negative. This is the case
of all graphs of class I obtained from a finite subgroup of $SU(N)$, 
since the two $M$ and $N$ algebras are then isomorphic to 
the class and the character algebras respectively
and their structure constants are thus integers in appropriate bases. 
This is also (by inspection) the case of the $A_n$, $E_6$ and $E_8$ 
Dynkin diagrams, as well as the $D_{2\ell}$ ones, if a suitable choice 
is made for the eigenvectors corresponding to the exponent $\Gl=2\ell-1$.
This is also the case of the fusion graphs of $\slh(N)$ at any 
level, since in that case the $M$ and $N$ algebras are both isomorphic 
to the fusion algebra, and the structure constants are the 
non negative fusion coefficients.

\noindent
Remark: A drawback of our approach is to rely on the choice  
of an explicit basis of the C-algebra(s), {\it i.e.} of the 
eigenvectors $\psi$. When some exponents have multiplicities, there are 
several choices of the $\psi$'s. Properties like the existence 
of the vertex 1, or the positivity of the $M$ and/or $N$
algebras depend on this choice. 
\bigskip
\subsec{3.2 Subalgebras} %
Suppose that the algebra defined by the structure constants $M$, 
the $M$ algebra in short, admits a sub-algebra in the following sense:

{\sl There is a subset $\hT$ of the set of exponents such that }
%
\eqn\IIIf{
\Gl, \Gm \in \hT \qquad M_{\Gl\Gm}^{\ \ \Gn}\ne 0 \Rightarrow 
\Gn\in \hT }
%
We  also {\it assume} that the set $\hT$ is 
stable under conjugation
 $\Gl \mapsto \bar \Gl$ and under the action of the $\IZ_N$ 
automorphism $\Gs$. Note that the stability under conjugation 
implies that the identity (denoted $\o$) belongs to the  set $\hT$. 
The stability under conjugation is itself satisfied if the $M$
algebra is a C-algebra (see the proof in [BI]). On the other hand, 
the assumption of stability under $\Gs$ is a non trivial condition, 
which eliminates some possibilities. For example in the case $E_6$, 
the subset $\{1,7\}$ of exponents defines a sub-algebra, but is not
stable under $\Gs:\ \Gl\mapsto 12-\Gl$. 

\medskip
We now make the further assumption that the subalgebra defined by $\hT$
has the following property 
\smallskip\nind
{\bf Property $\wp$} : {\sl There exists a partition of the set of
vertices $\CV =\bigcup_i T_i$ into 
classes such that 
\item{i)} if $\Gl\in \hT$, ${\psi^{(\Gl)}_a / \psi^{(\o)}_a }$ 
is independent of the representative $a$ within the class $T_i$;
\item{ii)} if $\Gl\notin \hT$, then for any class $T_i$, $\sum_{b\in T_i}
\psi^{(\Gl)}_b \psi^{(\o)}_b =0 $. }  
\smallskip\nind
Both conditions may be conveniently embodied in a single relation 
%
\eqn\IIIg{
 \forall \Gl, \forall T_i, \forall a\in T_i\qquad
 \sum_{b\in T_i} \psi^{(\Gl)}_b \psi^{(\o)}_b = \Gd_{\Gl\in \hT}
{\psi^{(\Gl)}_a \over \psi^{(\o)}_a }\sum_{b\in T_i} (\psi^{(\o)}_b )^2 }
%
Before looking at the implications of that property, let us first 
show that it is natural. One knows at least three situations in which 
it is realized.

\medskip
\nind {\bf Proposition 3.1} {\sl Property $\wp$ is verified in either of 
the three following cases
\item{1)} The subset $\hT$ is the $\Gs$ orbit of $\o$.  
\item{2)} The subset $\hT$ is associated with a $\IZ_r$
symmetry of the graph (in a sense to be explained).
\item{3)} The Pasquier algebra defines a C-algebra and the $M$ and $N$
structure constants are non negative. }
\smallskip 
First note that these cases are not exclusive of one another, as 
we shall see. Also, we have no claim that they exhaust all the situations
where property $\wp$  is satisfied. 
\nind 1) This is the most trivial case: the subset $\hT=\{\Gs^\ell(\o);
\ell=0, \cdots, N-1\}$ gives rise to a subalgebra, as follows from \Ir{b}:
if $\Gl,\Gm\in \hT$, $\Gl=\Gs^\ell(\o)$, $\Gm=\Gs^m(\o)$, thus
$$ M_{\Gl\Gm}^{\ \ \Gn}=\sum_a e^{{2i\pi\over N} (\ell+m)\Gt(a)}
\psi_a^{(\o)}\psi_a^{(\Gn)*} =\Gd_{ \Gn, \Gs^{\ell+m}(\o)}
$$
which is non vanishing only if $\Gn  \ \in \hT$. 
Then the classes $T_i$ contain all vertices with the same grading
$$ T_i= \{a\in \CV\ : \ \tau(a)=i\} \qquad i=1, \cdots, N \ . $$ 
Let us show that $\wp$ is fulfilled. Use the orthogonality of
the $\psi$'s  and relation \Ir{b}\ 
to write
\eqn\IIIh{
\eqalign{
\sum_b \psi_b^{(\Gl)} \psi_b^{(\Gs^\ell(\o))*} &= \Gd_{\Gl,\Gs^\ell(\o)}\cr
&= \sum_b \psi_b^{(\Gl)} \psi_b^{(\o)*} e^{-2i\pi{\ell\tau(b)\over N}} \cr
&=\sum_{T_j} \sum_{b\in T_j} 
\psi_b^{(\Gl)} \psi_b^{(\o)*} e^{-2i\pi{j\ell\over N}} \cr }}
%
Multiplying by $\exp 2i\pi {\ell J\over N}$ and summing over $\ell$ from 
$0$ to $N-1$ projects onto  $j=J$, giving
%
\eqn\IIIi{
\sum_{b\in T_J} \psi_b^{(\Gl)} \psi_b^{(\o)*} 
={1\over N} 
\sum_{\ell=0}^{N-1} e^{2i\pi{J\ell\over N}} \Gd_{\Gl,\Gs^\ell(\o)}\ .}
%
But  $\forall a\in T_J$, if $\Gl=\Gs^\ell(\o)$,
$ \psi_a^{(\Gl)}/\psi_a^{(\o)}= e^{2i\pi{J\ell\over N}}$. 
We also use the fact that 
$\sum_{b\in T_J} |\psi_b^{(\o)}|^2$ is independent of $J=0,1,\cdots, N-1$
and thus equal to $1/N$. This follows simply from the equality [PZ]
\eqn\IIIia{ \sum_{a\in T_J,\, b} \psi^{(\o)}_a (G_1)_{ab} \psi^{(\o)*}_b=
\Gc^{(\o)} \sum_{a\in T_J} |\psi^{(\o)}_a|^2= 
\Gc^{(\o)} \sum_{b\in T_{J+1}} |\psi^{(\o)}_b|^2 \ . }
Putting everything 
together, we  find that \IIIi\  amounts to relation \IIIg. 
\def\vp{\varphi}
\smallskip
\nind 2) Suppose that the graph admits a $\IZ_r$ automorphism, 
i.e. there a bijection between vertices $a \mapsto \vp(a)$, 
$\vp^r =id$, which preserves the grading $\tau(\vp(a))=
\tau(a)$ and the adjacency matrix $G_{\vp(a)\vp(b)}=G_{ab}$. 
One may then diagonalize $G$ (and all $G_p$) by means of eigenvectors
satisfying
%
\eqn\IIIk{
\psi^{(\Gl)}_{\vp(a)}= e^{2i\pi {q\over r}} \psi^{(\Gl)}_{a}}
%
i.e. of a given ``charge" $q=q(\Gl)\in \IZ_r$. 
Note that $\psi^{(\o)}$,  whose components are all positive, 
is necessarily of zero charge and that 
%
\eqn\IIIm{ M_{\Gl\Gm}^{\ \ \Gn} = e^{-2i\pi {q(\Gl)+q(\Gm)-q(\Gn)\over r}}
M_{\Gl\Gm}^{\ \ \Gn}}
does not vanish only if $q(\Gn)=q(\Gl)+q(\Gm)\ \mod N$. 
In particular a subalgebra of the 
$M$ algebra is provided by the set of exponents 
%
\eqn\IIIn{\hT= \{ \Gl | q(\Gl)=0 \}\ .}
%

\nind In that case, the classes of vertices $T_i$ are the $\vp$-orbits 
%
\eqn\IIIo{ \hbox{class of $a$}\, = T_i = \{\vp^j(a), j=0,\cdots, r-1\ .\}}
%
The length $\ell$ of the orbit may be smaller than $r$, it is 
a divisor of $r$; $\vp^\ell(a)=a$ implies that the only 
non vanishing $\psi^{(\Gl)}_a$ are those for which $q(\Gl)$ 
is a multiple of ${r\over \ell}$. One then computes easily,
if $a\in T_i$ belongs to an orbit of length $\ell$ of $\vp$
%
\eqn\IIIp{\eqalign{
\sum_{b\in T_i} \psi^{(\Gl)}_b \psi^{(\o)}_b &=\sum_{j=0}^{\ell -1}
\psi^{(\Gl)}_{\vp^j(a)}\psi^{(\o)}_{\vp^j(a)} \cr
&=\( \sum_{j=0}^{\ell-1} e^{2i\pi {q(\Gl)j\over r}} \)
{\psi^{(\Gl)}_a\over \psi^{(\o)}_a} \(\psi^{(\o)}_a \)^2\cr
&=\left\{\matrix{ 
 \ell{\displaystyle{\psi^{(\Gl)}_a\over \psi^{(\o)}_a}} 
\(\psi^{(\o)}_a \)^2
={\displaystyle{\psi^{(\Gl)}_a\over \psi^{(\o)}_a}}  \sum_{b\in T_i} 
\(\psi^{(\o)}_b \)^2
& \hbox{if} \ \ q(\Gl)=0\quad \mod\, r \cr 
0 \hfill & \hbox{ otherwise\ .} \cr
}\right.
\cr }}
%
This completes the proof that case 2) satisfies property $\wp$.

\nind 3) In the case of a C-algebra with non negative $M$
and $N$ (and thus $p$ 
and $q$) structure constants, one may apply the theorem of 
Bannai and Ito ([BI], theorem 9.9). $\hT$ defines a C-subalgebra. Then 
one defines the equivalence relation between exponents
$\Gm \sim \Gn$ if $\exists \Gl\in \hT\ :\ M_{\Gl\Gm}^{\ \ \Gn}\ne 0$.  
Thus there is a partition of the set $\Exp=\cup_\Ga \hT_\Ga$ into equivalence 
classes. One then proves that there exists a subset $T$ of the dual
set $\CV$ such that 
%
\eqn\IIIq{\forall a, \forall \hT_\Ga, \forall \Gl\in \hT_\Ga\qquad
 \sum_{\Gm\in \hT_\Ga} \psi^{(\Gm)}_a \psi^{(\Gm)}_1 = \Gd_{a\in T}
{\psi^{(\Gl)}_a \over \psi^{(\Gl)}_1 }\sum_{\Gm\in \hT_\Ga} 
(\psi^{(\Gm)}_1 )^2 
} 
%
(which is the dual form of \IIIg). This subset $T$ defines in turn 
a subalgebra of the dual $N$ algebra, in the sense that 
$a, b \in T,\ N_{ab}^{\ \ c}\ne 0 \Rightarrow c\in T$, and relation 
\IIIg\ is itself satisfied. 

\medskip
\nind{\it Consequences of property $\wp$}\par
\noindent 1. Let $\CN_i$ be the real positive numbers defined by
\eqn\IIIr{
\CN_i^{-2} = \sum_{a\in T_i} (\psi^{(\o)}_a)^2\ .}
Then for any $\Gl\in \hT$, define 
%
\eqn\IIIs{
\Psi_i^{(\Gl)}= \CN_i^{-1} {\psi^{(\Gl)}_a\over \psi^{(\o)}_a}}
%
By virtue of relation \IIIg, it is independent of $a \in T_i$ and 
in fact
%
\eqn\IIIt{
  \Psi_i^{(\Gl)}= \CN_i \sum_{b\in T_i} {\psi^{(\Gl)}_b \psi^{(\o)}_b}\ .}
%
\medskip
\nind {\bf Corollary 3.2}\quad {\sl 1. The  $\Psi_i^{(\Gl)}$, as $i$ runs over the 
classes $T_i$ and $\Gl$ over the subset $\hT$,   form an orthogonal
and complete system of rank equal to the cardinality of $\hT$. 
\eqna\IIIu
$$
\eqalignno{
\sum_{\Gl\in \hT}  \Psi^{(\Gl)}_i \Psi^{(\Gl)*}_j &= \Gd_{ij} &\IIIu a\cr
\sum_{i}  \Psi^{(\Gl)}_i \Psi^{(\Gm)*}_i &= \Gd_{\Gl\Gm}\quad \hbox{ for}
\ \ \Gl,\Gm\in \hT\ &\IIIu b\cr
}
$$
Thus the number of classes $T_i$ equals the
cardinality of $\hT$ (i.e. the dimension of the $M$ subalgebra). 
\nind 2. The subalgebra $M_{\Gl\Gm}^{\ \ \Gn}$, $\Gl,\Gm,\Gn\in \hT$, 
is diagonal in that basis 
%
$$ \eqalignno{ M_{\Gl\Gm}^{\ \ \Gn} 
 &= \sum_i {\Psi_i^{(\Gl)}\Psi_i^{(\Gm)}\Psi_i^{(\Gn)*}\over 
\Psi_i^{(\o)}} \ .   
& \IIIu c \cr}
 $$
} 
\nind {\it Proof }: the proof of the three statements \IIIu{a-c} 
follows the same line and makes use of \IIIg. Let us present it for 
\IIIu{b} only
\eqn\IIIw{\eqalign{ 
\sum_{i}  \Psi^{(\Gl)}_i \Psi^{(\Gm)*}_i &= 
\sum_i \CN_i^{-2}{\psi^{(\Gl)}_a \over \psi^{(\o)}_a}
{\psi^{(\Gm)*}_a \over \psi^{(\o)}_a} 
=  \sum_i 
{\psi^{(\Gl)}_a \over \psi^{(\o)}_a} 
\sum_{b\in T_i} {\psi^{(\Gm)*}_b \psi^{(\o)}_b}  \qquad \forall a \in T_i \cr
&=  \sum_i 
\sum_{b\in T_i} 
{\psi^{(\Gl)}_b \over \psi^{(\o)}_b} 
{\psi^{(\Gm)*}_b \psi^{(\o)}_b}  
= \sum_b {\psi^{(\Gl)}_b \psi^{(\Gm)*}_b}  = \Gd_{\Gl\Gm}\ . \cr }}


\bigskip
\noindent 2. We now  use the assumption of stability of $\hT$ under $\Gs$
to prove the
\medskip
\nind{\bf Corollary 3.3} {\sl All the vertices within a given class $T_i$
have the same grading $\Gt$.}
\smallskip
\nind {\it Proof} : Since $\o \in \hT$, let us evaluate relation \IIIg\ for 
$\Gl=\Gs(\o)$. 
\eqn\IIIx{
\sum_{b\in T_i} \psi^{(\Gs(\o))}_b\psi^{(\o)}_b 
= {\psi_a^{(\Gs(\o))}\over \psi_a^{(\o)}}
\sum_{b\in T_i}\( \psi^{(\o)}_b\)^2 }
thus, using \Ir{b}
\eqn\IIIy{
\sum_{b\in T_i}\( \psi^{(\o)}_b\)^2 e^{2i\pi {\Gt(b)-\Gt(a)\over N}}  
=\sum_{b\in T_i}\( \psi^{(\o)}_b\)^2} 
%
which is impossible if the phases are not equal to 0 ($\mod 2\pi$). 
(Geometrically the module of the l.h.s. is the end-to-end length
of a broken line, the r.h.s. the same when the line made of the same
segments is straight!). 

%
\bigskip
\bigskip
\subsec{3.3 Associated subgroup of the reflection group} %
\noindent We now prove that with any $M$ subalgebra satisfying the 
previous property $\wp$, we may associate a subgroup of the reflection 
group $\GC$. As before, let $T_i$ denote the classes in the
partition of $\CV$. 
\medskip
\nind {\bf Proposition 3.4} {\sl The operators $R_i=\prod_{a\in T_i} S_a$, 
$i=1, \cdots ,|\hT|$ generate a subgroup $\GC'$ of the reflection 
group $\GC$. $R_i$ is itself a reflection in the hyperplane 
orthogonal to 
\eqn\IVa{ \Gb_i= \CN_i \sum_{a\in T_i} \psi^{(\o)}_a \Ga_a\ , }
%
(with $\CN_i$ defined in \IIIr).}
\smallskip\nind 
Thus $\GC'$ is itself a reflection group in a subspace of dimension 
$|\hT|$ of the space $V$, generated by the $\Gb$. The partition 
into $T_i$ of the set of vertices provides the folding of the 
original ``Dynkin diagram": all the vertices $a\in T_i$ are 
mapped onto a single vertex $i$ of the folded diagram, with 
the new ``roots" $\Gb_i$ given by the real linear combinations \IVa. 
The new set of vertices inherits the $\Gt$ grading of the original
ones, thanks to Corollary 3.3.

\noindent
{\it Proof of Proposition}: By Corollary 3.3, all $a$ within  $T_i$ 
have the same grading $\Gt$ and thus the corresponding roots $\Ga_a$
are mutually orthogonal.  Consider the product $\prod_{a\in T_i} S_a$. 
Its action on an arbitrary vector of $V$ reads
%
\eqn\IVb{\prod_{a\in T_i} S_a x = x -\sum_{a\in T_i}\bra\Ga_a,x\ket \Ga_a\ .}
%
Compare it to the reflection $R_i$ in an hyperplane (through the origin)
orthogonal to the vector $\Gb_i= \sum_{a\in T_i} B_{ia} \Ga_a$, 
with real coefficients  $B_{ia}$. If $\Gb_i$ is normed to 
$\bra \Gb_i, \Gb_i\ket =2$, 
%
\eqn\IVc{\eqalign{
 R_i x&= x -\bra \Gb_i, x\ket \Gb_i \cr
&= x -\sum_{a,b\in T_i} B_{ia}B_{ib}\bra \Ga_b, x\ket \Ga_a\ . }}
%
$R_i$ and $\prod S_a$ have the same action in the subspace $\CE_i$ 
%
\eqn\IVd { \CE_i = \{x\ |\ \forall a \in T_i\quad  \sum_{b\in T_i} 
(\Ga_b,x) B_{ib}B_{ia}=(\Ga_a,x)  \} \ .}
%
The different $R_i$ act in a consistent way in the space $\CE= \cap_i \CE_i$
provided each $\Gb_i$ belongs to $\CE$, i.e.
%
\eqn\IVe{ \forall i,\, \forall a \in T_i,\, \forall j\quad
\sum_{b\in T_i \atop c\in T_j} B_{ia}B_{ib}B_{jc} g_{bc}= 
\sum_{c\in T_j} B_{jc} g_{ac}\ . } 
%
This condition is fulfilled by $B_{ia}= \CN_i \psi^{(\o)}_a$, i.e.
$\Gb_i$ as in Prop. 3.4. The normalisation factor ensures that 
$\bra \Gb_i, \Gb_i\ket =2$. To check eq. \IVe, one diagonalizes the 
metric $g_{ac}=
\sum_{\Gl\in \Expexp} g^{(\Gl)}\psi_a^{(\Gl)}\psi_c^{(\Gl)\, *}$. 
Then a short calculation shows that both sides of
\IVe\  are equal to 
$${ \sum_{\Gl\in \hT} g^{(\Gl)} \psi_a^{(\Gl)} 
{\psi_c^{(\Gl)\, *}\over \psi_c^{(\o)}}\ . } $$
This completes the proof of Prop. 3.4.

\bigskip

Note that the $\Psi_i ^{(\Gl)}$, $\Gl\in \hT$, defined in \IIIs\
 diagonalize the new metric:
%
\eqn\IVf {\eqalign{
g_{ij}&=(\Gb_i,\Gb_j)=\CN_i^{-1} \CN_j^{-1} \sum_{\Gl\in \hT}
g^{(\Gl)} {\psi^{(\Gl)}_a\over \psi^{(\o)}_a} 
{\psi^{(\Gl)\, *}_b\over \psi^{( \o)}_b}
\quad \forall \, a \in  T_i \ , \forall \, b \in  T_j 
 \cr 
&= \sum_{\Gl\in \hT}
g^{(\Gl)} \Psi^{(\Gl)}_i \Psi^{(\Gl)\, *}_j \ . \cr }} 
%
The geometry of the new root system $\{\Gb_i\}$ is encoded into
a graph which satisfies all axioms of sect. 1.2. From the
metric \IVf\ one derives the adjacency  matrices $G_{ij}=g_{ij} -2 
\Gd_{ij}$ and $(G_p)_{ij}$, 
since the color $\Gt(i)$ of each vertex is known (see \IIa, \Im\ and 
the remark at the end of sect 1.2). A determination (up to a
common factor) of the Coxeter number and exponents of the folded
graph is provided by 
\eqn\IVfa{h_{\hbox{{\sevenrm folded}}}  =h\ ,\qquad \Exp_{\hbox{\sevenrm
 folded}}  =\hT\ . }
%

\subsec{3.4 A guided tour of the zoo}

\nind There are a few cases which we may discuss in full generality. 
As we have seen (case 1 of Prop. 3.1), 
for any graph, the $M$ algebra admits a subalgebra 
associated with the exponents which are on the orbit of $\o$
%
\eqn\IVg{\hT=\{ \Gs^\ell(\o); \ell=0, \cdots, N-1\} \ .}
%
The graph may be folded in a trivial way, in which all vertices 
with the same color $\tau(a)=i$, $i=0,\cdots, N-1$ are 
mapped onto the same vertex $i+1$ of the folded graph.
 The scalar products of the new roots are also easy to compute.
For $i\ne j$: 
%
\eqn\IVh{
\eqalign{ \bra \Gb_i,\Gb_j \ket &= N \sum_{a: \Gt(a)=i-1\atop b: \Gt(b)=j-1}
\psi^{(\o)}_a \psi^{(\o)}_b \bra\Ga_a,\Ga_b\ket \cr
&= N \sum_{a: \Gt(a)=i-1} \psi^{(\o)}_a \sum_b (G_p)_{ab}\psi^{(\o)}_b \cr
&= N \sum_{a: \Gt(a)=i-1} (\psi^{(\o)}_a)^2\Gc_p^{(\o)}  \cr 
&= \Gc_p^{(\o)}  \cr }}
%
where $p= j-i \,\mod N\ne 0 \,\mod N$ 
and we have used again  
$\sum_{a: \Gt(a)=i} |\psi^{(\o)}_a|^2 
={1\over N}$.  
The resulting folded graphs are like the star-polygons mentioned 
above in the discussion of $\slh(N)_1$ (sect. 3.2), but with edges
now carrying the 
numbers $\Gc_p^{(\o)}$, {\it i.e.} $\Gc_p^{(0)}$ or $\Gc_p^{(\Gr)}$
of eq. \Ik.
In the case of $N=2$ class II graphs, (i.e. starting from an ordinary
$ADE$ graph), one obtains the Coxeter graph $I_2(h)$. For class I graphs, 
the graph $G$ has two vertices with $G_{12}=2$ edges 
between them, this is --up to the change of 
sign convention already mentioned in footnote 3-- the graph 
usually represented as $ \bullet \!\!{{\infty}\over{\qquad}}\!\circ$
and denoted in the literature  $\hA_1$ or $\tilde{A}_1$ or 
$A^{(1)}_1$ (we shall use the first notation). 
For $N>2$, the edges of the folded graphs encode scalar products 
which are (generically) larger than two, exposing the hyperbolic nature
of the geometry of the root system.

\fig{\rmn The folding of $\CA$  diagrams (top) into $\CB$ diagrams (bottom).
Left: $\slh(3)$ at level 3; right: $\slh(4)$ at level 2. 
Here for more clarity, we have replaced the notation
 ${{4}\over{\qquad}}$ ({\it i.e.} scalar products equal to
$\sqrt{2}$) by a dashed line. In the $\CB$ graphs: in the 
case of $\slh(3)_3$, the left  black dot represents the $\Gs$-orbit 
of $\Gl=(1,1)$, the right one $(2,2)$.  
In the case of $\slh(4)_2$, the left 
black dot represents the $\Gs^2$-orbit of $(1,1,1)$, the upper one
$(2,1,2)$. The right black square represents the orbit of
$(3,1,1)$, the upper one $(1,2,1)$.  }
{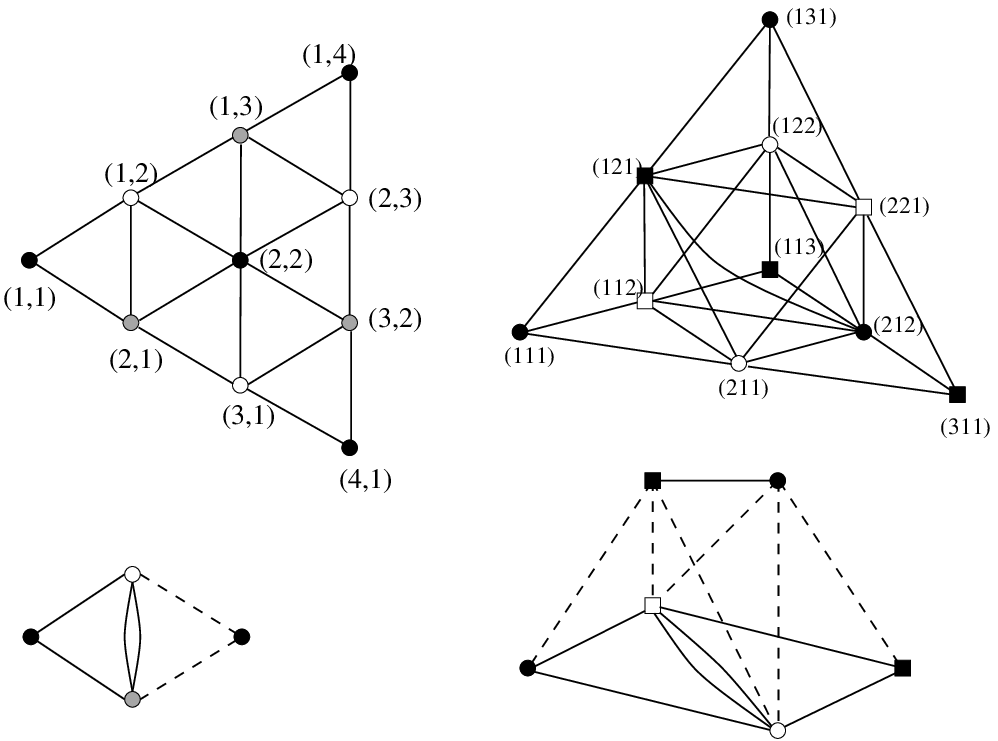}{90mm}\figlabel\folab

Another case of general validity is provided by the fusion graphs
of $\slh(N)_k$, when $k$ and $N$ are not coprimes $(N,k)=p$. Then 
it is easily seen that $\Gs^p$ is an automorphism of $\CP_{++}^{(h=N+k)}$
which preserves the grading $\tau$. It is thus a $\IZ_{N/p}$ 
automorphism in the sense of case 2) of Prop. 3.1 and 
may thus be used to fold the fusion graph. The result, which we 
denote $\CA^{(h)}\mapsto \CB^{(h)}$ by analogy with the conventional $A$ 
and $B$ Dynkin diagrams is illustrated in Fig. \folab\ in the two
cases of $\slh(3)_3$ and $\slh(4)_2$. Note that this {\it folding}
should not be confused with the {\it orbifolding} procedure 
$\CA^{(h)}\mapsto \CD^{(h)}$ of [Ko, FeG] (see Fig. \orbsutr); in 
the latter, the fixed points of $\Gs^p$ are taken with the 
multiplicity $N/p$ in the orbifold graph. These 
orbifold graphs may themselves be folded: see Fig.7 below. 

Finally we notice that for a  class I graph emanating from a finite
subgroup $H$ of $SU(N)$ (see sect 1.3), a subalgebra of the 
$M$ algebra is nothing else than a subalgebra of the {\it class}
algebra of $H$. Stability of $\Exp$ under $\Gs$ is equivalent 
to the stability of these classes under the $\IZ_N$ center, 
and property $\wp$ follows from the positivity of the 
structure constants of the class and character algebras. Thus
with any such subalgebra, we may associate a folding of the 
corresponding Dynkin diagram.

\medskip
We now review the folding of the $N=2$ graphs, {\it i.e.}
of the ordinary (simple or affine) Dynkin diagrams.

\fig{\rmn The folding of $\hA,\hD,\hE$ Dynkin diagrams. 
Classes $T_i$  encompass  vertices on the same vertical.
For the foldings $\hA_{2n-1}$ or $\hD_{n+2}$
 into $\hB_{n+1}$ or $\hC_n$, the parity of $n$ is irrelevant, 
{\nineit i.e.} the color of the rightmost vertices could be reversed. }
{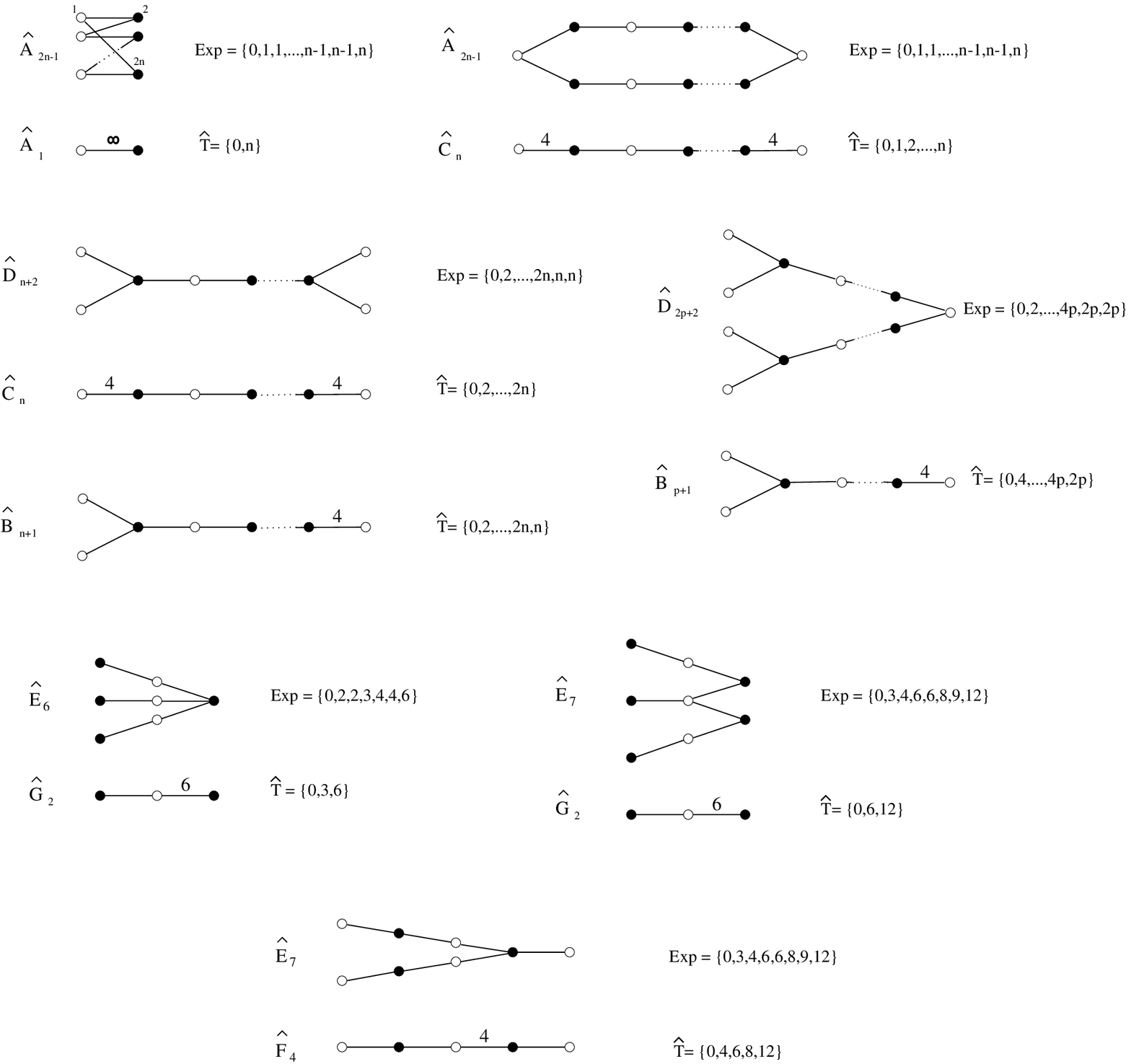}{150mm}\figlabel\foldhat

For class I, one may fold any simply laced affine 
Dynkin diagram onto $\hA_1$ as just explained. In Fig. \foldhat\ this is 
displayed only for the folding of $\hA_{2n-1}$ into $\hA_1$.
All the other foldings exploit an obvious symmetry of one of 
the diagrams of type $\hA$, $\hD$, $\hE_6$ or $\hE_7$.  
One may fold $\hD_{n+2}$  into $\hB_{n+1}$
or into $\hC_n$, and  $\hD_{2p+2}$ into $\hB_{p+1}$. 
Likewise, one may fold $\hE_7$ into $\hF_4$
or into  $\hG_2$. The latter may also be obtained from $\hE_6$. 
Note that in order to discuss the folding of $\hA_{2n-1}$ into $\hC_{n+1}$
or of $\hD_{n+2}$ into $\hB_{n+1}$, one cannot use the natural 
basis $\psi$ associated with characters of the corresponding 
subgroups of $SU(2)$, but one must choose another one in which 
positivity of the $M$ and $N$ algebras is lost but the existence 
of subalgebras is manifest.
These cases thus escape the case 3) of Prop. 
3.1 but are still covered by case 2).

\fig{\rmn The folding of $ADE$ Dynkin diagrams. 
Classes $T_i$  encompass  vertices on the same vertical.}
{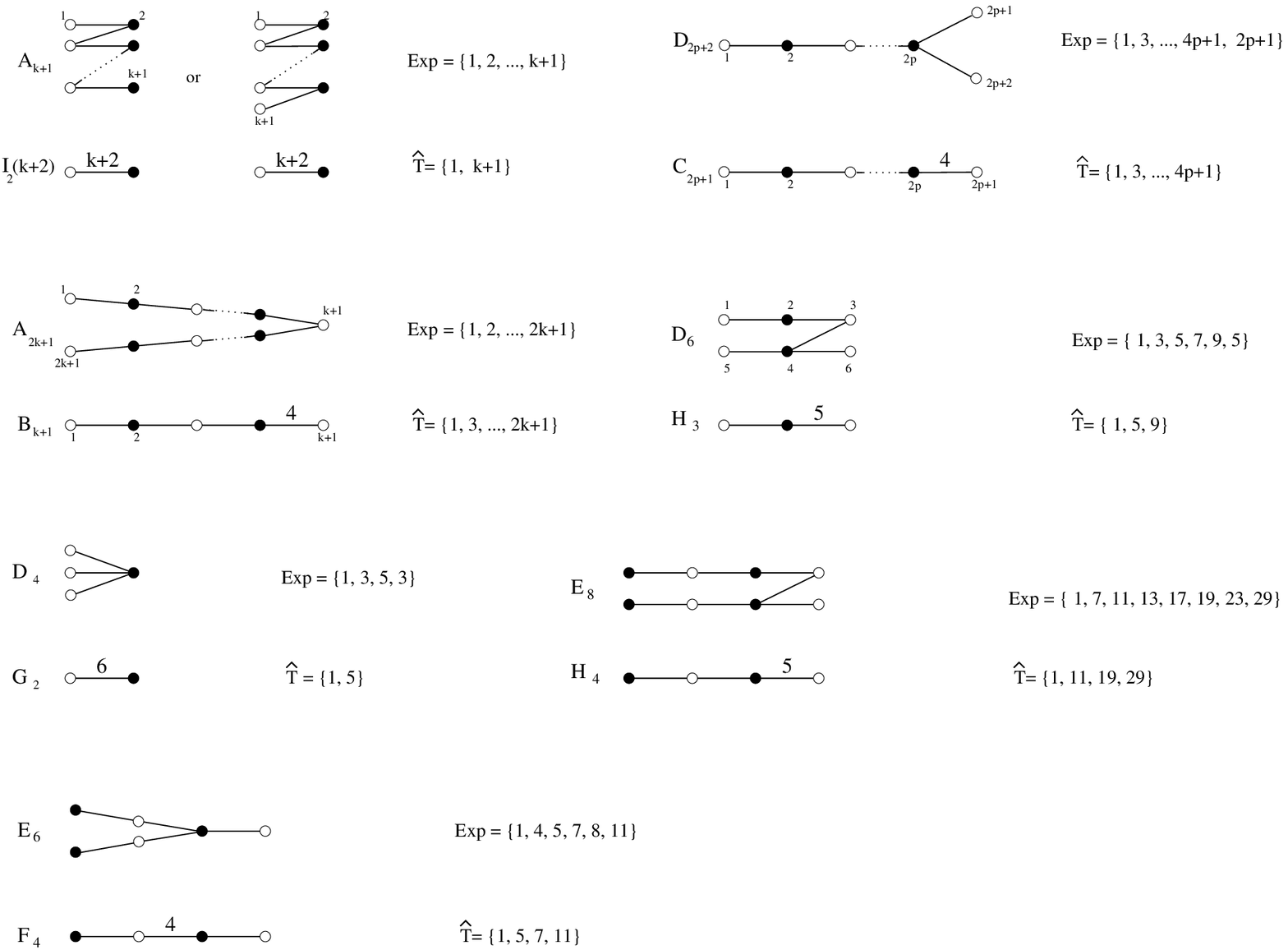}{135mm}\figlabel\folsim

For class II (fig \folsim), {\it i.e.} ordinary simple 
Dynkin diagrams, the situation is quite parallel. Any simply 
laced diagram with Coxeter number $h$ may be folded into $I_2(h)$. 
This is shown in Fig. \folsim\ only for the $A_n$ graphs and for $D_4\mapsto
I_2(6)\equiv G_2$. Some  other foldings result from some obvious
symmetry: $A_{2k+1}\mapsto B_{k+1}$, $D_{2p+2}\mapsto C_{2p+1}$, 
$E_6 \mapsto F_4$ (and also $D_4 \mapsto G_2$). But the  cases
of $D_6\mapsto H_3$ or of $E_8 \mapsto H_4$ require more ingenuity 
if one proceeds empirically. The virtue of the procedure discussed in 
the present paper is to yield these cases as well as the others as resulting 
from the discussion of sect. 3.2 and 3.3, namely from Prop. 3.1. 
In fact all cases but  $D_{2p+2}\mapsto C_{2p+1}$ fall into case 3) of 
that Proposition. For the latter, as discussed before, to 
expose the $M$-subalgebra, one has to choose a basis in which 
positivity of the $M$ and $N$ is lost. 

In these $N=2$ cases, where there is an {\it a priori}
classification of all possible graphs 
with a positive definite or semi-definite metric $g$ [BH], 
one may verify that  
the folding procedure based on $M$-subalgebras is exhaustive: 
it produces all non simply laced diagrams 
starting with the simply laced ones. It would be desirable
to have such a result in general.

\fig{\rmn A sample of some $N=3$ diagrams and their folding.
{\ninebf (a)} the graph $\CD^{(6)}$ of fig. \orbsutr\
and three of its possible foldings; same 
convention as in fig. 4 for the dashed lines, 
the dotted lines stand for ${{6}\over{\qquad}}$ ({\nineit i.e.} a 
scalar product equal to $\sqrt{3}$).
{\ninebf (b)} a graph of Coxeter number $h=8$, its set 
of exponents and a folding: the exponents of the subset 
$\hat T$ are underlined;
{\ninebf (c)} a graph of Coxeter number $h=12$, its set 
of exponents, a subset $\hT$ underlined leads to the folding
on the right; 
{\ninebf (d)} folding of the graph of fig. \subgr(b). }
{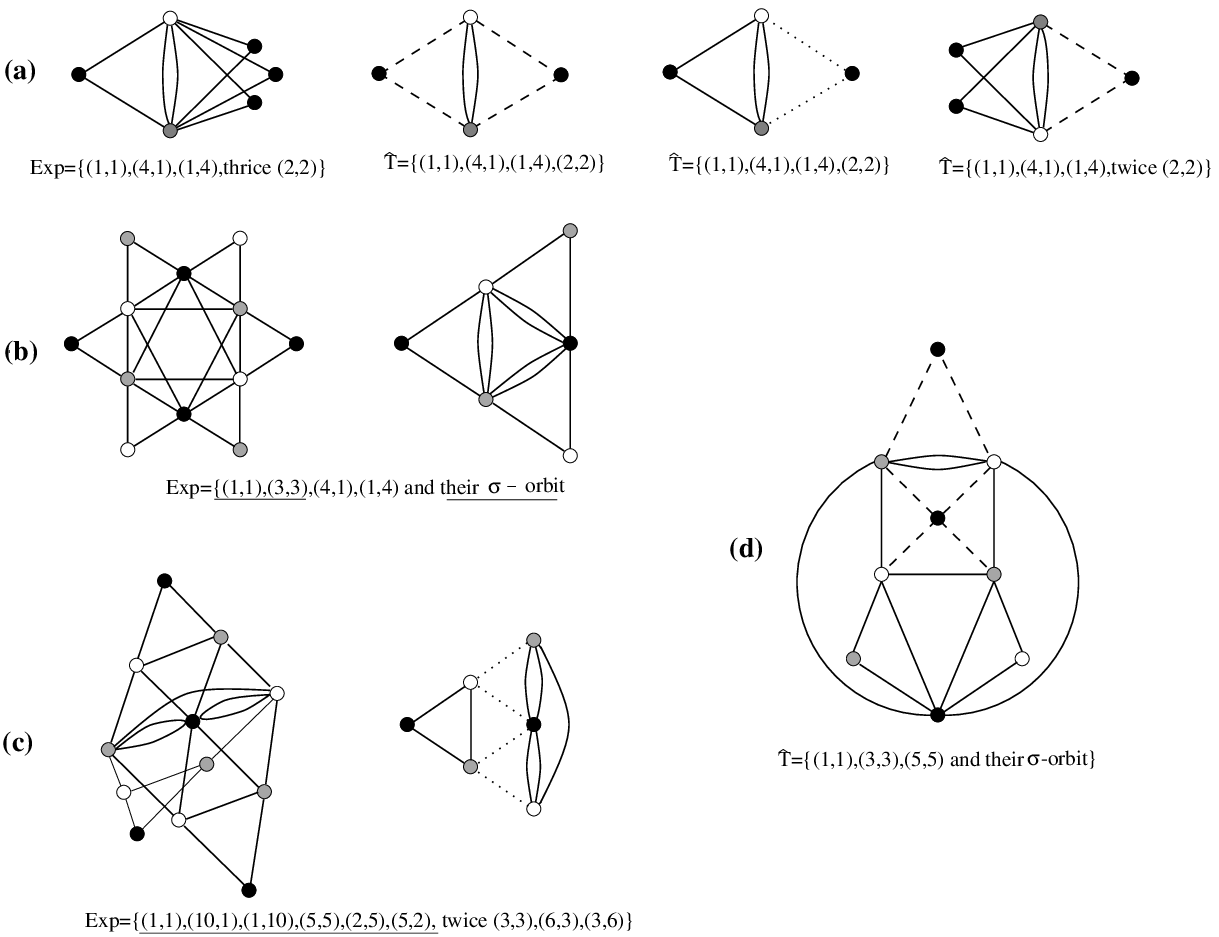}{135mm}\figlabel\folsutr

\medskip
As for the folding   of $N>2$ graphs, we content ourselves 
with a sample of graphs and foldings in Fig. \folsutr. 



%
\vglue8mm
\penalty -5000
\AHead{4. Relations with singularity theory}
\penalty 10000
%

It is well known that generalized Dynkin diagrams may be 
associated with singularities. In particular the $ADE$ diagrams are
associated  with simple singularities [AGZV]. A subclass of the 
diagrams studied in this paper may be given such an interpretation. 

This should only be true for (some) graphs of class II. 
For a polynomial singularity  $W=0$, one looks at the monodromy 
of homology cycles of a non-singular level set $\{x\in \IC^m,\, 
|x|< \Gd\,|\,  W(x)=\Ge\}$ 
 as $\Ge$ circles around the origin [AGZV]. 
When $W$ is quasi-homogeneous, this 
monodromy operator is necessarily of finite order. 
If the putative Coxeter element $R$ discussed in sect. 2.2 
 is to be identified with this monodromy operator of the singularity, 
as is the case in many explicit cases, then  graphs of class I 
for which $R$ has been 
proved to be of infinite order (Prop. 2.3)
are excluded from this interpretation. 

The case that is best understood is the class of ``fusion 
graphs" of $\slh(N)_k$. As explained in sect. 1.3, these graphs encode
the fusion of integrable representations of $\slh(N)_k$
by the fundamental representations. The fusion algebra of 
$\slh(N)_k$ has been proved by Gepner [Ge] to be isomorphic 
to the quotient algebra 
$\IC[X_1,\cdots, X_{N-1}]/\CI(\partial_i\tilde W_{N,k})$, 
quotient of the algebra of polynomials in $N-1$ variables by 
the ideal generated by the first partial derivatives of the 
``fusion potential"
\eqn\Va{\tilde W_{N,k}(X_1,\cdots,X_{N-1})=
\Big({d\over dt}\Big)^{k+N} 
\hbox{log}( 1-t X_1 +\cdots +(-t)^{N-1} X_{N-1} +(-t)^N)|_{t=0}\ .}
It is then very natural to consider the quasi-homogeneous part of this
polynomial and to conjecture that the fusion graphs 
of the present paper are the Dynkin diagrams associated  with 
the singularity 
\eqn\Vb{ W_{N,k}(X_1,\cdots,X_{N-1})=
\Big({d\over dt}\Big)^{k+N} 
\hbox{log}( 1-t X_1 +\cdots +(-t)^{N-1} X_{N-1} )|_{t=0}\ .}
This conjecture [Z] has now been proved by Gusein-Zade and 
Varchenko [GZV]. This proof has also the following interesting consequences:
\item{-} All the isomorphisms painfully discussed in a case-by-case analysis
in [Z] are obtained in one shot! For example, the group associated with 
the graph $\CA^{(5)}$ of fig.\fussutr\ is isomorphic to the ordinary 
$D_6$ finite Coxeter group since both describe the singularity 
$W=XY^2+X^5$.  Likewise the $\CA^{(6)}$ graph describes the 
same singularity $W=X^6+aX^2Y^2+Y^3$ 
as the graph denoted $J_{10}$ in [AGZV].
\item{-} In all these isomorphisms of pairs of graphs/groups, the 
two  root systems may be obtained from one another 
by a sequence of sign changes $\Ga_a\to -\Ga_a$ and braiding : 
$\{\Ga_1, \cdots, \Ga_{a-1},\Ga_a, \cdots , \Ga_n\}
\to \{\Ga_1, \cdots, \Ga_{a},S_a \Ga_{a-1}, \cdots , \Ga_n\}$. 
\item{-} The monodromy operator of the singularity is the Coxeter 
element defined at the end of sect. 2.2 by the lexicographic ordering.
\item{-} Rank-level duality : the reflection groups  associated 
with the fusion graphs of $\slh(N+1)_k$ and $\slh(k+1)_N$ are isomorphic. 
This too was conjectured in [Z] and proved in [GZV].
\foot{\rmn 
Beware! this is not what is usually called the rank-level duality, 
which rather connects $sl(N)_k$ and $sl(k)_N$. There is, however, 
a relation between the two concepts [NS].}

\medskip
In many cases, however, it
is known that there is  no description of the graph or group 
in terms of a single polynomial singularity. This is 
evidenced by looking at  the characteristic  polynomial
of the monodromy operator. Suppose 
we have a singular polynomial $W(X_1,\cdots, X_m)$, 
  quasihomogeneous of degree 1 with degree$(X_\ell)=d_\ell$, and suppose
the eigenvalues of its monodromy 
operator are written as $(-1)^{m-1} \exp 2i\pi \zeta_j $, then one 
proves [Va] that
\eqn\Vc{\CP(t):=\sum_{j=1}^n t^{\zeta_j}= \prod_{\ell=1}^m {t^{d_\ell}-t \over
1-t^{d_\ell}}  = t^{\Sigma d_\ell}  \prod_{\ell=1}^m {1- t^{1-d_\ell} \over
1-t^{d_\ell}} \ .}
The last product in \Vc\ is the Poincar\'e polynomial (in some
fractional power of $t$) counting the 
elements in the local ring of the singularity. 
For the Coxeter element $R$ of sect. 2.2, we may compute
$\CP(t)$ knowing the $\zeta_\Gl=-{(e_N,\Gl)\over h}$. 
If $R$ is to be interpreted as 
the monodromy operator of  a  quasi-homogeneous singularity, then 
according to \Vc, 
$\CP(t)$ must have its zeros on the circle $|t|=1$. 
Now, it is easy to check that in several cases discussed in the present 
paper, the operator $R$ does not meet this condition.  
An example is provided 
by the graph $\CD^{(6)}$ of fig. \orbsutr\ (alias $D_4^{(1,1)}$ of [S])
for which 
$$ \CP(t) = (1+ 4\sqrt{t} +t)\ ,  $$
the roots of which lie clearly out of the unit circle. 

\medskip
Finally we note that whenever a graph may be associated with a singularity, 
its various foldings are associated with boundary singularities, as 
discussed in [Y,Sh,AGZV]. 

\bigskip
All these connections with singularity theory have a natural 
transcription in ``physical" language, in the context of
two-dimensional $\CN=2$ superconformal field theories and/or
 topological field theories. We shall only sketch the line of arguments.  
It is believed that there exist $\CN=2$ superconformal 
 and topological field theories associated with each of the graphs discussed 
here.
Cases associated with a singularity are said to have 
a Landau-Ginsburg (L-G) potential. Not all theories possess
such a L-G potential. In all cases, however, as already mentioned  
above, Cecotti and Vafa [CV]  have found that in the study of
the so-called $tt^*$ equations, one encounters a monodromy 
operator with very specific properties. 
It is suggested that the Coxeter element of this paper 
is the monodromy operator of Cecotti and Vafa. 
On the other hand,  Dubrovin has developped a geometric
interpretation of the axioms of two-dimensional 
topological field theories (tft),  in terms of Frobenius manifolds. 
There too, a differential system which relates the flat coordinates
of two independent flat metrics plays a central role, and its monodromy group
is --at least in some cases-- generated by reflections. There
are many indications that the reflection group of the present paper
describe this monodromy group (or a subgroup of it) in the case
of topological theories based on $sl(N)$. The challenge 
would be to reconstruct the topological field theory, at least in genus zero
-- a solution of the Witten-Dijkgraaf-Verlinde-Verlinde equations--, 
from the graph. This presumably involves the study of the space of
orbits of the group, in the spirit of [D].
In that context, going from a graph to a folded graph must produce another
topological theory, obtained by restricting the moduli space of the 
original one.  This is what happens in the simplest cases 
of the minimal tft's [D,Zu].

\AHead{5. Conclusion and perspectives} 
\secno=5
 The present paper has considered a class of graphs which generalize in a 
natural way the ordinary Dynkin diagrams and thus lead to reflection 
groups. These graphs satisfy a certain number of axioms. 
Admittedly, these axioms lack elegance and conciseness! A more
compact and conceptually simpler way of encapsulating them would be 
welcome. 

We have shown that many solutions may be obtained in a systematic way
by folding. Folding of graphs is usually found using their symmetries. 
Roughly speaking, these symmetries of the vertex set imply restrictions
in the dual (``Fourier") exponent set, consistent with their algebra
($M$ algebra). We have found that conversely and more generally, 
such restrictions --with additional  technical assumptions-- may lead
to new foldings not associated with an obvious symmetry. 

As already pointed out, our analysis in terms of $C$-algebras and
subalgebras relies on explicit choices of basis in these algebras
({\it i.e.} explicit choices of the eigenvectors $\psi$). It would 
be desirable to have a more intrinsic and geometric 
standpoint. The same applies to property $\wp$, which appears as 
a useful technical assumption, whose intrinsic meaning remains
however obscure. 

Another direction which deserves more work is the connection with 
singularity theory. In particular, in cases not associated with a 
quasihomogeneous polynomial, is there a substitute? 

Finally the connections with superconformal field theories 
and topological field theories reviewed at the end of 
previous section are still awaiting a thorough discussion.

\penalty 10000
       
\vbox{
\AHead{ 
Acknowledgements} 
I want to thank the Taniguchi Foundation 
and  the organisors of this Symposium, K. Saito, M. Kashiwara, 
A. Matsuo and I. Satake
for providing exceptional conditions of  work and of scientific
exchanges in a beautiful location. I benefited a lot 
from discussions with many participants, especially 
K. Saito, I. Satake and P. Slodowy. I would like also to acknowledge the 
stimulating criticisms and suggestions of M. Bauer, P. Di Francesco, 
V. Petkova and  A. Varchenko during the preparation of this paper.}

\bigskip
\RefHead{References}

\itemitem{[AGZV]} V.I. Arnold,  and S.M. Gusein-Zade, A.N. Varchenko,
{\it Singularities of differentiable maps}, Birk\"auser, Basel 1985.

\itemitem{[BH]}
N. Bourbaki, {\it Groupes et Alg\`ebres de Lie}, chap. 4--6, Masson 1981; 
J.E. Humphreys, {\it Reflection Groups and Coxeter Groups}, 
Cambridge Univ. Pr. 1990. 

\itemitem{[BI]} E. Bannai, T. Ito, {\it Algebraic Combinatorics I: Association
Schemes}, Benjamin/Cummings (1984).

\itemitem{[BLM]} S. Berman, Y.S. Lee and R.V. Moody, 
{\it J. Algebra} {\bf 121} (1989) 339-357.

\itemitem{[C]} 
H.S.M. Coxeter, {\it Ann. Math.} {\bf 35} (1934) 588-621.

\itemitem{[CV]} S. Cecotti and C. Vafa, 
{\it Nucl.Phys.}{\bf 367} (1991) 359-461;
{\it Comm. Math. Phys.}{\bf 158} (1993) 569-644
.

\itemitem{[D]} B. Dubrovin,
{\it Nucl. Phys.} {\bf B 379} (1992) 627-689:
{\tt hep-th/9303152};
Springer Lect. Notes in Math. {\bf 1620} (1996) 120-348: 
{\tt hep-th/9407018};  B. Dubrovin and Y. Zhang,  
{\it Extended affine Weyl groups and Frobenius manifolds}, {\tt hep-th/9611200}.

\itemitem{[DFZ1]} P. Di Francesco and J.-B. Zuber, 
{\it Nucl. Phys.} {\bf B338} (1990) 602-646. 

\itemitem{[DFZ2]}
P. Di Francesco and J.-B. Zuber, 
in {\it Recent Developments in Conformal Field Theories}, Trieste
Conference, 1989, S. Randjbar-Daemi, E. Sezgin and J.-B. Zuber eds., 
World Scientific 1990; 
 P. Di Francesco, {\it Int.J.Mod.Phys.} {\bf A7}  (1992) 407-500.

\itemitem{[FFK]} W.M. Fairbairn, T. Fulton and W.H. Klink, 
{\it J. Math. Phys} {\bf 5} (1964) 1038-1051.

\itemitem{[FG]} P. Fendley and P. Ginsparg, {\it Nucl. Phys. }{\bf B324} 
(1989) 549-580.

\itemitem{[FH]}{W. Fulton and J. Harris, {\it Representation Theory}, 
Springer Verlag, ex 2.37 page 25, corr P 517. }

\itemitem{[G]}
D. Gepner, {\it Comm. Math. Phys.} {\bf 141} (1991) 381-411.

\itemitem{[GHJ]} F. Goodman, P. de la Harpe and V.F.R. Jones, 
{\it Coxeter graphs and towers of algebras}, {\bf 14} MSRI Publ., 
Springer (1989).

\itemitem{[GZV]} S.M. Gusein-Zade and A. Varchenko, 
{\it Verlinde algebras and the intersection form on vanishing cycles}, 
{\tt hep-th/9610058}.

\itemitem{[K]} B. Kostant, Soci\'et\'e Math\'ematique de France, 
Ast\'erisque (1988) 209-255.

\itemitem{[Ko]} I.K. Kostov,  Nucl. Phys.  {\bf B 300} [FS22] (1988)
559-587. 

\itemitem{[KP]} V.G. Kac and D.H. Petersen {\it Adv. Math.} {\bf 53}
(1984) 125-263. 

\itemitem{[MK]} J. MacKay, {\it Proc. Symp. Pure Math.}
{\bf 37} (1980 183-186.

\itemitem{[MP]} 
R.V. Moody and J. Patera, {J.Phys.A} {\bf 26} (1993) 2829-2853.

\itemitem{[NS]} S.G. Naculich and H.J. Schnitzer, 
{\it Superconformal coset equivalence from level-rank duality}, 
{\tt hep-th/9705149}. 

\itemitem{[O]} 
{A. Ocneanu, communication at the Workshop
{\it Low Dimensional Topology, Statistical 
Mechanics and Quantum Field Theory},
 Fields Institute, Waterloo, Ontario, April 26--30, 1995.}

\itemitem{[P]} V. Pasquier,  J.Phys. {\bf A20} 5707-5717 (1987); 
Th\`ese d'Etat, Orsay, 1988.

\itemitem{[Pe]} V.B. Petkova, private communication and to appear. 

\itemitem{[PZ]} V.B. Petkova and J.-B. Zuber, 
{\it Nucl. Phys. B} {\bf B463} (1996) 161-193: {\tt hep-th/9510175}; 
{Conformal Field Theory and Graphs}, {\tt hep-th/9701103}.  

\itemitem{[S]} K. Saito, 
{\it Publ. RIMS, Kyoto Univ.}
{\bf 21} (1985) 75-179, {\bf 26} (1990) 15-78; 
K. Saito and T. Takebayashi,  preprint RIMS-1089.  

\itemitem{[Sh]} O.P. Shcherbak, {\it Russ. Math. Surveys} 
{\bf 43:3}  (1988) 149-194.

\itemitem{[St]} J. Steenbrink, {\it Compositio Math.} {\bf 34} 
(1977) 211-223.

\itemitem{[Va]} A.N. Varchenko, {\it Inv. Math.} {\bf 37} (1976) 253-262.

\itemitem{[Ve]} E. Verlinde, Nucl. Phys. {\bf B300} [FS22] (1988) 360-376.

\itemitem{[Y]} J. Sekiguchi and T. Yano, 
{\it Sci. Rep. Saitama Univ.} {\bf IX} (1980) 33-44; 
T. Yano, 
{\it ibid.} 61-70.

\itemitem{[Z]} J.-B. Zuber, {\it Comm. Math. Phys.} {\bf 179} (1996) 
265-294.

\itemitem{[Zi]} J.-B. Zuber, {Conformal, Integrable and Topological 
Theories, Graphs and Coxeter Groups}, in {\it XIth International Congress of 
Mathematical Physics}, Paris July 1994, 
D. Iagolnitzer edr, International Press 1995, p 674-689,
{\tt hep-th/9412202}. 

\itemitem{[Zr]} J.-B. Zuber, {\it C-algebras and their applications 
to reflection groups and conformal field theories}, 
proceedings of the RIMS Symposium, Kyoto, 16-19 December 1996,
{\tt hep-th/9707034}.

\itemitem{[Zu]} J.-B. Zuber, {\it Mod. Phys. Lett} {\bf A8} (1994) 749-760.

\bigskip

\affil{\rm Jean-Bernard Zuber \hfill \break
Service de Physique Th\'eorique, \hfill \break
 CEA Saclay, F-91191 Gif sur Yvette Cedex, (France)\hfill\break
email:  zuber@spht.saclay.cea.fr}

\bye